\documentclass[paper]{JHEP3}
\usepackage{epsfig,bbm}
\usepackage{simplewick}
\newcommand{\be}{\begin{equation}} % only untightened
\newcommand{\ee}{\end{equation}}
\newcommand{\bea}{\begin{eqnarray}} % only untightened
\newcommand{\eea}{\end{eqnarray}}
\newcommand{\nn}{\nonumber}
\newcommand{\bmp}{\noindent\begin{minipage}{16cm}}
\newcommand{\emp}{\end{minipage}\vskip 7mm} % 7mm untightened
\def\lsim{\mathrel{\raise.3ex\hbox{$<$\kern-.75em\lower1ex\hbox{$\sim$}}}}
\def\gsim{\mathrel{\raise.3ex\hbox{$>$\kern-.75em\lower1ex\hbox{$\sim$}}}}
\newcommand{\ksl}{\mathbin{k\mkern-10mu\big/}}
\newcommand{\psl}{\mathbin{p\mkern-10mu\big/}}
\newcommand{\intron}[1]{}%{#1}
\newcommand{\hs}{\hat s}

\newcommand{\mf}{\hat m_f}

\def\ra{{\rightarrow}}

\def\sfrac#1#2{{\textstyle\frac{#1}{#2}}}

\title{Superweakly interacting dark matter from the Minimal Walking Technicolor}

\author{Kimmo Kainulainen\footnote{kimmo.kainulainen@jyu.fi} ,
        Jussi Virkaj\"arvi\footnote{jussi.virkajarvi@jyu.fi}\
        \\
        \\ Department of Physics, P.O.Box 35 (YFL), 
        \\ FI-40014 University of Jyv\"askyl\"a, Finland, 
        \\ and 
  	    \\ Helsinki Institute of Physics, P.O.~Box 64, 
  	    \\ FI-00014 University of Helsinki, Finland.\\}	    
\author{Kimmo Tuominen\footnote{kimmo.tuominen@jyu.fi}\,\,\footnote{On leave of absence from Department of physics, University of Jyv\"askyl\"a} \\
{CP}$^{ \bf 3}${-Origins}, 
Campusvej 55,\\ DK-5230 Odense M, Denmark\footnote{{ C}entre of Excellence for { P}article { P}hysics { P}henomenology devoted to the understanding of the {Origins} of Mass in the universe} \\and\\
Helsinki Institute of Physics, 
P.O.Box 64, \\ FI-000140, University of Helsinki, Finland}

\abstract{We study a superweakly interacting dark matter particle motivated by minimal walking technicolor theories. Our WIMP is a mixture of a sterile state and a state with the charges of a standard model fourth family neutrino. We show that the model can give the right amount of dark matter over a range of the WIMP mass and mixing angle. We compute bounds on the model parameters from the current accelerator data including the oblique corrections to the precision electroweak parameters, as well as from cryogenic experiments, Super-Kamiokande and from the IceCube experiment. We show that consistent dark matter solutions exist which satisfy all current constraints. However, almost the entire parameter range of the model lies within the the combined reach of the next generation experiments.}

\keywords{dark matter theory, dark matter experiments, physics of the early universe}

\preprint{CP3-Origins-2009-26}
\begin{document}
\section{Introduction}

Our universe appears to be to a high precision flat and dominated by dark matter (DM) and dark energy components of an unknown origin. The case for dark matter is particularly strong, with several lines of evidence pointing to its existence from galactic rotation curves to CMB and colliding galaxy clusters~\cite{Clowe:2006eq}. The exact value of the present day DM abundance to some degree depends on the cosmological model for the accelerated expansion. In the current standard model, where the expansion history is attributed to a cosmological constant, one finds the best fit values $\Omega_{\rm m,0} \simeq 0.20$ and $\Omega_{\Lambda} \simeq 0.76$ respectively~\cite{Dunkley:2008ie}. We shall use these values throughout this paper as a reference, although we keep in mind that in other models for the acclerated expansion considerably different values of DM density are predicted~\cite{Mattsson:2007tj}.

The WIMP (Weakly Interacting Massive Particle) paradigm asserts that DM consists of particles which interact very weakly under the Standard Model charges and whose presence is felt only through their gravitational interactions. In addition to giving consistent explanation to a host of independent cosmological  observations, the WIMP paradigm is appealing because good WIMP candidates arise naturally in the context of several well motivated theories beyond the Standard Model, such as supersymmetry~\cite{Lahanas:2006mr}.  Another popular extension of the Standard Model is Technicolor \cite{Hill:2002ap}. In TC the mass patterns of the standard model gauge bosons are explained by new strong dynamics without the need to invoke  fundamental scalar particles. The masses of elementary fermions are then typically explained by the extended technicolor (ETC) interactions~\cite{Hill:2002ap,Dimopoulos:1979es,Eichten:1979ah}. An interesting alternative is provided by hybrid models featuring in addition to composite TC scalars also fundamental scalar particle which couples to the matter fields via Yukawa interactions \cite{Simmons:1988fu,Kagan:1991gh,Carone:1992rh,Antola:2009wq}. 

Simple TC models are severely constrained by the limits on flavor changing neutral currents and by the appearance of unwanted additional light pseudo-Goldstone bosons. These problems are avoided in {\em walking technicolor} theories where the technicolor gauge coupling evolves slowly due to a near-conformal behavior.  Achieving conformal behavior using only fundamental representations for technifermions requires a large number of techniflavours, and this leads to a large oblique $S$ parameter $S\approx {\cal O}(1)$ contrary to the observed value $S\approx 0$. However, considering higher fermion representations increases the screening of matter fields and allows approaching conformality with much smaller particle content~\cite{Lane:1989ej,Corrigan:1979xf}. The phase diagrams for higher representations relevant for walking technicolor model building were constructed and candidates for minimal models of walking technicolor were proposed in \cite{Sannino:2004qp,Dietrich:2005jn}. In particular, it was shown that with just two techniflavors in the two-index symmetric (i.e. the adjoint) representation of the SU$_{\rm{TC}}$(2) gauge group the theory is already close to, or even within, the conformal window; this two-color and two-flavor theory is called the Minimal Walking Technicolor (MWT) model. In \cite{Hong:2004td,Dietrich:2005jn,Dietrich:2005wk} it was shown that MWT is compatible with precision measurements and that the MWT composite Higgs boson can be light, with mass of the order of few hundred GeV \cite{Dietrich:2005wk}. Collider phenomenology of MWT model has been studied in detail e.g. in~\cite{Foadi:2007ue,Foadi:2008ci,Antipin:2009ch,Antipin:2009ks,Frandsen:2009fs}, and recently the underlying strong dynamics has been investigated on the lattice~\cite{Catterall:2007yx,Catterall:2008qk,Hietanen:2008mr,Hietanen:2009az,DelDebbio:2009fd}. 

Despite the simple matter content of MWT model, it possesses rich dynamics and implies interesting phenomenological consequences. In particular, the model provides possible DM candidates which are the objects of interest for us here. Both technibaryonic~\cite{Foadi:2008qv} and leptonic~\cite{Kainulainen:2006wq} candidates have been investigated earlier. In this paper our goal is to update and extend the latter study where the WIMP was identified with a state with the quantum numbers of a standard fourth generation neutrino, whose existence is required by the internal consistency of the MWT model. However, it was observed already in~\cite{Enqvist:1988we}, that such particle interacts too strongly and thereby gives rise to a too small DM-density to be consistent with the observations in a universe with an ordinary expansion history. In~\cite{Kainulainen:2006wq} we overcame this problem by proposing that the universe were expanding faster than normal during the dark matter freeze-out because of a dynamical dark energy dominance. However, the improved constraints from cryogenic dark matter searches have subsequently ruled out this scenario~\cite{Angle:2008we}. In ref.~\cite{Kainulainen:2006wq} we only considered pure Majorana and pure Dirac neutrinos, while more complicated mass mixing structures are in fact naturally generated in the MWT-context \cite{Antipin:2009ks}. Given the most general mass matrix for the neutrino sector, it is easy to arrange that the (lightest) state corresponding to the dark matter has its SM charges further suppressed by a small mixing angle $\theta$. This suppression helps in two ways over the scanario studied in ref.~\cite{Kainulainen:2006wq}. First, the reduced annihilation cross section leads to a larger relic density, allowing agreement with the observed DM-density even within the standard expansion history of the universe. Second, the reduced couplings lessen the strength of the WIMP-nucleon interactions, allowing the model to evade the current cryogenic constraints. The stability of our WIMP can be simply imposed by giving a new discrete quantum number to all new fields relevant for the DM-sector (or a new $Z_2$-symmetry), in analogy to the usual $R$-parity in the case of supersymmetric theories. For earlier studies related to very heavy neutrinos, see e.g. \cite{Fargion:1994me}.

The idea of a superweakly interacting fourth family neutrino WIMPs was already considered in ref.~\cite{Enqvist:1990yz} and its relevance for the MWT-scenario was suggested in~\cite{Kainulainen:2006wq}, after which the MWT-WIMP scenario including the mixing was considered in ref.~\cite{Kouvaris:2007iq}. Here we improve and extend these analysis in several respects. First, we present a more complete TC-implementation of the effective WIMP-scalar interaction sector and couplings. Second, we compute the WIMP cross sections much more accurately; where refs.~\cite{Enqvist:1990yz,Kouvaris:2007iq} used a simple leading term longitudinal WW-approximation for the gauge boson final states, we compute the complete annihilation cross sections including all gauge and Higgs boson final states. Our final results depend both quantitatively and qualitatively on using the complete cross sections. Third, we carefully compute constraints on the mass spectrum from the latest cryogenic searches and from the oblique corrections implied by the high precision electroweak data.  Finally we identify an important and to our knowledge previously unnoticed effect of relative Majorana phases on the predictions of the model. These phases have to be introduced to ensure that the Majorana mass eigenstates are positive definite, and to span the entire mass matrix parameter space, one needs to supplement the physical masses and mixing angles with the signs of bilinear phase factor products. These products enter to WIMP-Higgs and WIMP-$Z$ couplings and then lead to different predictions for the relic density and for the oblique parameters as well as to different sensitivity on cryogenic searches. In the present case only one nontrivial phase product exists and we give all our results for the resulting two independent scenarios. Our final results is that MWT-WIMP is a consistent dark matter particle over a range of parameters forming a narrow twisting band in the WIMP mass -- mixing angle ($m_2,\theta$)-plane, roughly falling within the region $\sin \theta <0.4$ and 25GeV $\lsim m_2 \lsim$ 45 GeV or $m_2 \gsim$ 80-100 GeV. Most of this parameter space will be accessible to the combination of the ongoing or the next generation cryogenic experiments and neutrino detectors. In addition, the mass of the heavier neutral state $m_1$ and the mass of the charged lepton $m_E$ completing the weak lepton doublet, are constrained by the high precision electroweak data such that only narrow finite strips in the $(m_2,m_E)$-plane are allowed for a given cosmologically acceptable ($m_2,\theta$)-solution. 

The paper is organized as follows: we will introduce the underlying technicolor model and discuss the precision electroweak constraints in section \ref{sec:model}. In section \ref{sec:omega} we describe and give our main results of the computation of the relic abundance $\Omega_N$ as a function of the mass of the lightest neutrino state, $m_2$ and the neutrino mixing angle $\sin\theta$. Then, in section \ref{sec:cryo} we consider the constraints to the model by applying the bounds from the accelerators, cryogenic searches and from neutrino detectors such as Super-Kamiokande and IceCube, including estimates for the future reach of these experiments. Our conclusions and outlook are presented in section \ref{sec:checkout}. Finally,  in the appendix we present the relevant details of the cross section calculations. 

%%%%%%%%%%%%%%%%%%%%%%%%%%%%%%%%%%%%%%%%%%%%%%%%%%%%%%%%%%%%%%%%%%%%%%%%%%%%%%%
%
% SECTION 1  TC
%
%%%%%%%%%%%%%%%%%%%%%%%%%%%%%%%%%%%%%%%%%%%%%%%%%%%%%%%%%%%%%%%%%%%%%%%%%%%%%%%

\section{Minimal Walking Technicolor and the Fourth Generation of Leptons}
\label{sec:model}

\subsection{Model Lagrangian and mass terms}

In the MWT model to be studied here, the electroweak symmetry breaking is driven by the gauge dynamics of two Dirac fermions in the adjoint representation of SU$_{\rm{TC}}$(2) gauge theory. The key feature of this model is that it is (quasi) conformal with just one doublet of technifermions \cite{Sannino:2004qp}. However, since the technicolor representation is three dimensional, the number of weak doublets is odd and hence anomalous \cite{Witten:fp}. A simple way to cure this anomaly is to introduce one new weak doublet, singlet under technicolor and QCD color \cite{Sannino:2004qp,Dietrich:2005jn} in order not to spoil the walking behavior and to keep the contributions to the oblique corrections as small as possible. Hence, the model requires the existence of a fourth generation of leptons. The anomaly free hypercharge assignments for the new degrees of freedom have been presented in detail in \cite{Dietrich:2005jn}. Here we simply note that there exists an assignment which makes the techniquarks and the new lepton doublet appear exactly as a regular standard model family from the weak interactions point of view and it is this assignment that we will consider throughout this work.  

The technicolor sector of the theory confines at electroweak scale and is described better in terms of a chiral effective theory than using the fundamental techniquark and -gluon degrees of freedom. The global symmetry breaking pattern is SU$(4)\rightarrow {\rm SO}(4)$, with nine goldstone bosons. Three of these are absorbed into the longitudinal degrees of freedom of the weak gauge bosons, and the low energy spectrum is expected to contain six quasi Goldstone bosons which receive mass through extended technicolor interactions~\cite{Hill:2002ap,Appelquist:2002me,Appelquist:2004ai}. Their phenomenology has been investigated elsewhere \cite{Foadi:2007ue,Foadi:2008ci}. Here we will use the result obtained in \cite{Foadi:2007ue} that the conservation of the hypercharge allows only the effective SM-like scalar Higgs to couple to the fermions. Hence, we re-introduce a (composite) Higgs doublet into the theory. We denote the left-handed fourth generation lepton doublet by $L=(N_L, E_L)$ and the right-handed SU$_L(2)$ singlets as $E_R$ and $N_R$. To take into account the effects of the scalar sector on these leptons up to and including dimension five operators, we consider following effective interactions
\begin{eqnarray}
{\mathcal{L}}^{I}_{{\rm{Mass}}} 
  &=& (y \bar{L}_L H E_R+ {\rm{h.c.}})+C_D\bar{L}_L\tilde{H}N_{R}
\nonumber \\
  &+& \frac{C_{L}}{\Lambda}(\bar{L}^c\tilde{H})(\tilde{H}^TL)
   +  \frac{C_{R}}{\Lambda}(H^\dagger H)\bar{N}^c_{R}N_{R} 
   +{\rm{h.c.}}
\label{scalar_fermion}
\end{eqnarray}
where $\tilde{H}=i\tau^2H^\ast$ and $\Lambda$ is a suppression factor related to the more complete (ETC) ultraviolet theory. The first terms in Eq.~(\ref{scalar_fermion}) lead to the usual (Dirac) mass for the charged fourth generation lepton, and the remaining terms allow for more general mass structure of the fourth neutrino. After symmetry breaking the effective Lagrangian (\ref{scalar_fermion}) gives rise to a neutrino mass term:
\begin{equation}  
   -\frac{1}{2}\bar{n}_L^c 
   \left(\begin{array}{cc} M_L & m_D \\ m_D &  M_R\end{array}\right) n_L
   + h.c. \,,
\label{eq:massmatrix}
\end{equation}
where $n_L=(N_{L}, N_{R}^{~c})^T$, $m_D=C_Dv/\sqrt{2}$ and $M_{L,R}=C_{{L,R}}v^2/2\Lambda$, where $v$ is the vacuum expectation value of the effective Higgs field. The special cases are a pure Dirac and a pure Majorana neutrino which are obtained, respectively, by discarding dimension five operators and by removing the right handed field $N_{R}$. The most general mass matrix contains, even after the field redefinitions, one complex phase. However, in this paper we shall restrict ourselves to the case of real mass matrix. Mass eigenstates are two Majorana neutrinos which are related to the gauge eigenstates by a transformation 
\begin{equation}
  N = O n_L + \rho O^T n_L^c \,,
\label{Neigen}
\end{equation}
where $N \equiv (N_1,N_2)^T$ and $O$ is an orthogonal $2\times 2$ rotation matrix, where the associated mixing angle is 
\begin{equation}
  \tan 2\theta = \frac{2m_D}{M_R-M_L} \,.
\label{eq:tan}
\end{equation}
The phase-rotation matrix $\rho = {\rm diag}(\rho_1,\rho_2)$ is included above to ensure that the physical masses $m_{1,2}$ are positive definite. Indeed, the eigenvalues of the mass matrix in (\ref{eq:massmatrix}) are
\begin{equation}
  \lambda_{\pm}=\frac{1}{2}\Big(M_L+M_R\pm\sqrt{(M_L-M_R)^2+4m_D^2} \;\Big) \,.
\label{eq:masseigen}
\end{equation}
Because the signs and relative magnitudes of $M_{L,R}$ and $m_D$ are arbitrary, the eigenvalues $\lambda_\pm$ can be either positive or negative. However, choosing  independent phases as $\rho_{\pm}={\rm{sgn(}}\lambda_{\pm})$ we get positive $m_\pm=|\lambda_\pm|$ as required. For our purposes it will be convenient to express everything in terms of the physical mass eigenvalues $m_1 > m_2$ and the mixing angle $\sin\theta$ instead of the Lagrangian parameters $M_L$, $M_R$ and $m_D$. While working with physical parameters has obvious advantages but the downside is that the connection between the physical and the Lagrangian parameters is not always straightforward. 

\subsection{Accounting for the mixing phases}
\label{sect:rho-business}
Let us now discuss in detail the role of the phase-rotation matrix $\rho$ introduced in the previous subsection. There are two distinct but equivalent ways to treat these phases. First, one may embed them into the definition of the mass eigenstates which leads to unitary rotation matrix in Eq. (\ref{Neigen}). Then phases appear explicitly in the interaction vertices and related Feynman rules while the Majorana eigenstates satisfy $n^c=n$, making the field operator $\rho$-independent. This formulation was considered for this model in \cite{Antipin:2009ks}. The second posibility, which we shall adopt here, is to retain the definition (\ref{Neigen}) with an orthogonal rotation. This has the consequence that the Majorana eigenstates satisfy $n^c=\rho n$ i.e. the phases now appear explicitly in the field operators corresponding to the Majorana eigenstates. Then the phases have to be properly taken into account in the contractions and propagators when evaluating various processes. This issue will be considered to some detail in the appendix.

\FIGURE[t]{\epsfig{file=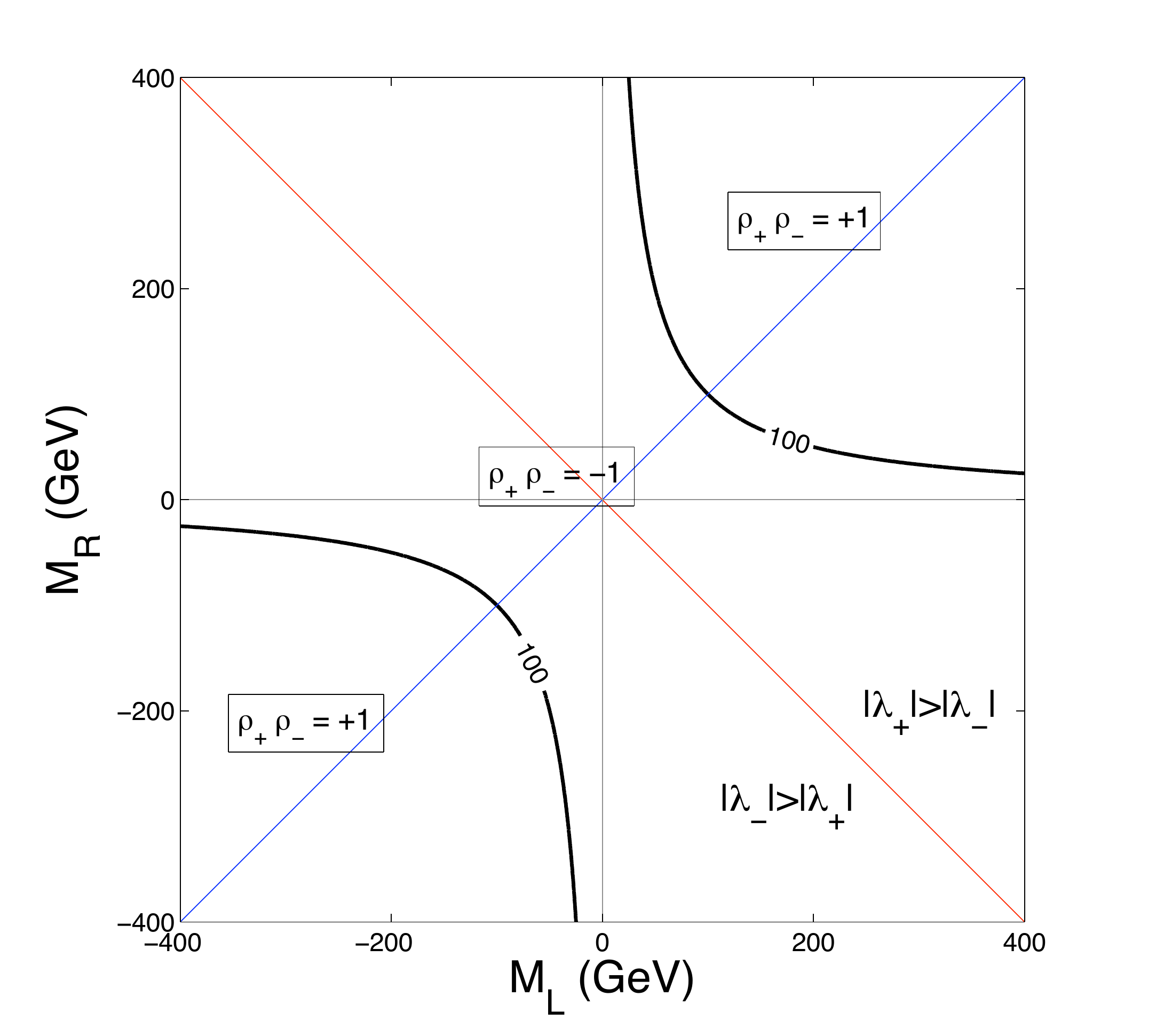,width=9.2cm}
\caption{Shown is a constant-$m_D$ slice of the mass parameter space, which is divided into physically disjoint regions by the sign of the effective phase $\rho_+\rho_-$. The physical parameters are invariant under the mirroring symmetry with respect to the line $M_R=-M_L$.}
\label{fig:st1}  }
Let us now describe the structure of the parameter space and the role of the phases. In Fig. \ref{fig:st1} we show the parameter space along a plane of some nonzero value of $m_D$.  Let us first suppose that $M_R+M_L\ge 0$.  Then the eigenvalue $\lambda_+$  is (\ref{eq:masseigen}) is always positive and larger than $|\lambda_-|$. We choose to denote by $N_1$ the heavier state, so that $\rho_1 \equiv \rho_+ = 1$ and our WIMP is always associated with $N_2$ with the mass $m_2 = m_-$. The sign of $\rho_2 = \rho_-$ is defined by the ratio of $m_D^2$ and $M_LM_R$: $\rho_2  = -1$ if $m^2_D>M_LM_R$, and otherwise $\rho_2=+1$. Individual phases are not observable, but relative phase corresponding to the product of the two is:
\begin{equation}
\rho_1\rho_2 = \rho_+\rho_- \equiv \rho_{12} \,.
\end{equation}
In particular this phase will show up explicitly in the various couplings of the mass eigenstates. The situation with respect to the product phase is again illustrated in Fig.~\ref{fig:st1}: $\rho_+\rho_-$ is positive in the upper right corner separated by the black solid line and negative elsewhere in the region $M_R+M_L\ge 0$. From Eq.~(\ref{eq:tan}) we see  that $\tan 2\theta$ becomes infinite along the the line $M_L=M_R$, corresponding to a maximal mixing $|\theta |\equiv \pi/4$. We define the mixing angle to be zero in the limit $M_R-M_L>>|m_D|$, whereby $0 \le |\theta | \le \pi/4$ below and $\pi/4 \le |\theta | \le \pi/2$ above the line of maximal mixing. The sign of the mixing angle is determined by the sign of $m_D$. Physically, in the region below the line of maximal mixing, our WIMP is predominantly a sterile right chiral state while in the region above the line it is predominantly a left chiral state with ordinary weak interaction strength. It is thus clear that the area of most interest for us is the rightmost quadrant bounded by the diagonal lines in the phase space in Fig.~\ref{fig:st1}.

Now consider the case $M_R+M_L \le 0$. In this area we have that $\lambda_-$ is always negative and $|\lambda_-| > |\lambda_+|$. Thus, in this region we have to associate our heavier state (which we always choose to label with $N_1$) with $N_-$, while the WIMP is always the now lighter $N_2 \equiv N_+$. Here we have always $\rho_1 = \rho_- = -1$ while the sign $\rho_1=\rho_+$ again depends on the relative magnitude of $M_LM_R$ and $m_D^2$; the resulting division to distinct areas according to the sign of $\rho_{12} =\rho_-\rho_+$ is again shown in Fig.~~\ref{fig:st1}. There is an obvious symmetry in the phase space about the reflection along the line $M_L+M_R = 0$. Indeed, all that happens in this reflection is that the eigenvalues $\lambda_\pm$ and their associated eigenstates exchange roles, but all physical parameters $\theta$, $m_1$, $m_2$ and $\rho_{12}$ remain invariant. That is, the reflection corresponds to a mere relabeling $1\leftrightarrow 2$ as the two regions can be mapped to each others by a redefinition of the phases of the states. We shall thus concentrate on the region $M_R+M_L \ge 0$ from now on. Nevertheless, for each triplet of physical mass and mixing parameters $m_1$, $m_2$ and $\theta$ the original parameter space contains two physically distinct solutions labelled by the relative phase $\rho_{12}$. In what follows, we shall always give the results for both possibilities.

We conclude this subsection by noting that all the typical special cases are contained within the $\rho_{12} = -1$ portion of the parameter space: the pure left- or right-handed Majorana states correspond to $M_R=0$ and $M_L=0$ axes in the plane $m_D=0$, respectively, while the $M_L=0$ and $M_R=0$ planes for nonzero $m_D$ correspond to usual seesaw scenarios. Finally the Dirac limit corresponding to the axis $M_L=M_R=0$ is also contained only in the $\rho_{12}=-1$ domain. It is perhaps due to this reason that the other domain with $\rho_{12}=+1$ has so far gone unnoticed in the literature.

%%%%%%%%%%%%%%%%%%%%%%%%%%%%%%%%%%%%%%%%%%%%%%%%%%%%%%%%%%%%%%%%%%%%%%%%%%%%%%%
% Subsection Couplings
%%%%%%%%%%%%%%%%%%%%%%%%%%%%%%%%%%%%%%%%%%%%%%%%%%%%%%%%%%%%%%%%%%%%%%%%%%%%%%%%

\subsection{Couplings}
\label{sec:couplings}

For the analysis of the relic density we need the couplings of the neutrino mass eigenstates to the weak gauge bosons and to the Higgs boson. These are easily found out by applying the appropriate phase- and rotation transformations defined in the previous section. We shall write down only the terms relevant for our calculations.  For the $Z$ and $W^\pm$ bosons we find that
\begin{eqnarray}
W^+_\mu\bar{N}_L\gamma^\mu E_L 
  &=& \sin\theta \; W^+_\mu \bar{N}_{2L}\gamma^\mu E_L+\cdots  \nonumber\\
Z_\mu\bar{N}_L\gamma^\mu N_L 
  &=& \sin^2\theta \; Z_\mu \bar{N}_{2L}\gamma^\mu N_{2L} \nonumber \\
  &+& \sfrac{1}{2}\sin 2\theta \;  Z_\mu \, ( \bar{N}_{1L}\gamma^\mu N_{2L}                              
                             + \bar{N}_{2L}\gamma^\mu N_{1L} ) + \cdots  \,, 
\end{eqnarray}
where the omitted terms contain interactions of the heavy $N_1$ field only. These couplings are diagonal in the mixing and therefore do not involve the phase factor $\rho_{12}$. However, neutral current involves mixing and these couplings do depend on $\rho_{12}$. One finds:
\begin{equation}
\bar{N}_2\gamma^\mu Z_\mu P_L N_1+\bar{N}_1\gamma^\mu Z_\mu P_LN_2 = \bar{N}_2(\beta+\alpha\gamma_5)\gamma^\mu Z_\mu N_1,
\label{eq:neutralcurrent}
\end{equation}
where 
\begin{equation}
\alpha=\sfrac{1}{2}(1+\rho_{12}) \quad {\rm  and} \quad \beta=\sfrac{1}{2}(1-\rho_{12}) \,.
\label{eq:alphabeta}
\end{equation}
Thus, for $\rho_{12} = -1$ the neutral current interaction of our WIMP is purely axial vector and for $\rho_{12} = +1$ purely vector. Usually in the literature dealing with the interactions of Majorana neutrinos, only the first possibility is mentioned. 

The effective interaction terms involving the Higgs and the lighter neutrino eigenstate are
\begin{eqnarray}
{\mathcal{L}}_{NH}  &=& 
\frac{gm_2}{2M_W}\Big( \; C_{22}^h h\bar{N_2}N_2+C_{21}^h h\bar{N_1}(\alpha-\beta\gamma^5)N_2 
\nonumber\\ 
&& \phantom{Ham} + C_{22}^{h^2} h^2\bar{N_2}N_2 \,\Big)+
\frac{m_H^2}{2v}h^3 + \cdots \,,
\label{higgs_interactions}
\end{eqnarray}
where we have again omitted the interaction terms which do not contain $N_2$ and hence are not needed in our analysis. The factors $\alpha$ and $\beta$ are  defined in Eq. (\ref{eq:alphabeta}) and the factors $C^h_{22}$, $C^h_{21}$ and $C^{h^2}_{22}$ are defined in the left panel of Table~\ref{c_table}.  \\

\TABLE[t]{
\begin{tabular}{| l | c | c |}
\hline
 & Scenario I & Scenario II \\
\hline
$C_{22}^h$     & $1-\frac{1}{4}\sin^22\theta\, R_{-}$ 
               & $\sin^2\theta$ \\ [2mm]
$C_{21}^h$     & $-\frac{1}{4}\rho_{12}\sin4\theta\,R_-$ 
               & $ \frac{1}{2}\rho_{12}\sin2\theta\,R_+$    \\[2mm]
$C_{22}^{h^2}$ & $\frac{1}{2}-\frac{1}{4}\sin^22\theta\, R_{-}$ 
               & $\frac{1}{2}\sin^2\theta (1-\cos^2\theta R_-) $ \\ [1mm]
\hline
\end{tabular}

\caption{ Coefficients of the Lagrangian (\ref{higgs_interactions}) for two the distinct mass generating scenarios described by Eqs.~(\ref{scalar_fermion}) and (\ref{scalar_fermion2}). We have defined $R_{\pm}\equiv 1 \pm \rho_{12}\frac{m_1}{m_2}$. }
\label{c_table}
}

So far our model building has rested on the assumption that the (composite) Higgs is the sole source of elementary fermion masses. However, this is not necessarily the case, and there may be other (composite or even fundamental) scalars whose condensation leads to mass terms for the matter fields. To illustrate such possibilities, we consider as an alternative to the model Lag\-rangian~(\ref{scalar_fermion}), the case where the right-handed neutrino mass originates from a Standard Model singlet scalar field $S$. 
%which is singlet under the Standard Model charges:
%
\begin{eqnarray}
{\mathcal{L}}^{II}_{{\rm{Mass}}} 
  &=& (y \bar{L}_L H E_R+ {\rm{h.c.}})+C_D\bar{L}_L\tilde{H}N_{R}
\nonumber \\
  &+& \frac{C_{L}}{\Lambda}(\bar{L}^c\tilde{H})(\tilde{H}^TL)
   +  C_{R}S\bar{N}^c_{R}N_{R} 
   +{\rm{h.c.}}
\label{scalar_fermion2}
\end{eqnarray}
This model is similar to the usual see-saw neutrino mass generation mechanism, although here the singlet $S$ does not need to be a fundamental scalar. To specify the model completely one should give a potential for $S$. However, none of the parameters of this potential are needed in our analysis; we may assume that the vacuum expectation value for $S$ is generated through interactions with the Higgs, i.e. we do not need additional sources of spontaneous symmetry breaking for the dynamics of the $S$-field. In what follows, we will refer to the scenario with just only the doublet Higgs field as Scenario I and to the case with Higgs and a singlet scalar as Scenario II. The interactions between the Higgs and the neutrino can be generically described by the Lagrangian (\ref{higgs_interactions}) for both scenarios. The precise form of the coefficients is given in Table~\ref{c_table}.

Let us conclude this section by noting that to keep the WIMP stable, we have to exclude all renormalizable operators of the form
\begin{equation}
W^+_\mu \bar N_{iL} \gamma^\mu \ell_R \,, \quad 
Z_\mu \bar \nu_{\ell L} \gamma^\mu N_R \,, \quad 
\bar L_L H \ell_R \,, \;...
\end{equation}
which would mix the new WIMP-sector to the other standard model fields $L_{\ell L}$, $\ell_R$, or $\nu_{\ell R}$ where $\ell = e, \, \mu, \, \tau$. The simplest way to achieve this is to assume that all new fields relevant for the DM-sector, $L_L$, $E_R$ and $N_R$ share a new discrete conserved quantum number. Then all allowed Lagrangian terms must obey a new $Z_2$-symmetry under the exchange of these new fields. This construction is analogous to the $R$-parity imposed in Minimal Supersymmetric Standard Model to guarantee stability of protons (and consequently of the LSP).

\subsection{Oblique constraints}
\label{sec:oblique}

The fourth generation of leptons is constrained by current accelerator data. From LEP we know that the charged lepton $E$ has to be more massive than the $Z$ boson and if the fourth generation neutrino has standard model interaction strength, it needs to be heavier than $M_Z/2$ in order to evade the constraint from $Z$-pole observables. In the case of neutrino mixing considered in this work, the lighter state can have a substantial right-handed component and hence interact only very weakly. This, as we shall see, can allow this state to escape the LEP bounds even when its mass is less than $M_Z/2$. In addition to these direct bounds, the parameters of the fourth generation leptons are constrained by oblique corrections, i.e. due to their contribution to the vacuum polarizations of the electroweak gauge bosons. These contributions are conveniently represented by the $S$ and $T$ parameters \cite{Peskin:1990zt}.   

The oblique corrections in MWT model with the general mass and mixing patterns considered here have been studied in detail in \cite{Antipin:2009ks}. We also note that there exists two extensive fits performed by the LEP Electroweak Working Group (LEPEWWG) \cite{:2005ema} and independently by the PDG \cite{Amsler:2008zzb}. Both fits find that the SM, defined to lie at $(S,T)=(0,0)$ with $m_t=170.9$ GeV and $m_H=117$ GeV, is within $1\sigma$ of the central value of the fit. The two fits disagree on the central best-fit value: LEPEWWG finds a central value $(S,T)=(0.04,0.08)$ while including the low energy data the PDG  finds $(S,T)=(-0.04,0.02)$.  Since the actual level of coincidence inferred from these fits depends on the precise nature of the fit, we allow a broader range of $S$ and $T$ values, roughly corresponding to the $3\sigma$ contour. Concretely we require $0\le S\le 0.2$ and $0\le T\le 0.5$. From the results of \cite{Antipin:2009ks} it can be inferred that these values can be accommodated easily within the parameter space of the leptonic sector.

In this work we will supplement the constraints on the parameter space by requiring in addition to the saturation of the precision constraints that the lightest neutrino mass eigenstate provides the correct relic density to match the observed DM abundance. Hence, we adopt the following strategy: We will first determine the mass of the lightest eigenstate, $m_2$ and the corresponding mixing angle $\sin\theta$ such that the relic abundance $\Omega_{N_2}(m_2,\sin\theta)\sim 0.2$. This constraint between $m_2$ and $\sin\theta$ will be practically independent of the values of the remaining mass parameters $m_1$ and $m_E$ in the leptonic sector provided $m_2\ll m_1,m_E$. Then, for each $m_2$ and $\sin\theta$ constrained through $\Omega_{N_2}$ we will determine the allowed values of $m_1$ and $m_E$ by requiring that resulting contribution to $S$ and $T$ is within the bounds quoted above.
We will therefore now move to describe the evaluation of the relic density and the constraints from earth-based direct dark matter searches and return to the oblique corrections and constraints in Sec. \ref{sec:resultsoblique}.

\section{Relic Density}
\label{sec:omega}

The relic abundance $\Omega_{N_2}$ is computed in the standard way. We start from the Lee-Weinberg equation for the scaled WIMP number density~\cite{Lee:1977ua}:
\begin{equation}
\frac{\partial f(x)}{\partial x} 
        = \frac{\langle v\sigma\rangle 
         m_2^3 x^2}{H} (f^2(x)-f_{eq}^2(x)) \,, 
\label{ecosmo1}
\end{equation}
where $m_2$ is the WIMP mass and we have introduced the variables
\begin{equation}
  f(x) \equiv \frac{n(x)}{s_E}, \quad {\rm and} \quad     
  x \equiv \frac{s_E^{1/3}}{m_2},                                                          
\end{equation}
where $s_E(T)$ is the thermal entropy density 
at the temperature $T$. Given $s_E$, the Hubble parameter $H(T) = (8\pi\rho(T)/3M_{\rm Pl}^2)^{1/2}$ and the average WIMP annihilation rate $\langle v\sigma \rangle$, Eq.~(\ref{ecosmo1}) is easily solved numerically. Here we assume that the expansion of the universe follows the standard adiabatic expansion law, so that $H$ and $s_E$ can be computed from their standard thermal integral expressions. Typically our WIMPs are freezing out at $T\sim {\cal O}(1-10)\rm \; GeV$, so that the uncertainties in $s_E$ arising from the QCD phase transition play no role for us. After the present ratio of $N_2$-number-density to the entropy density $f(0)$ is found from Eqn.~(\ref{ecosmo1}), the fractional density parameter $\Omega_{N_2}$ of the Majorana WIMPs becomes
\begin{equation}
\Omega_{N_2}\simeq 5.04 \times 10^5 m_2 f(0)\,.
\label{ecosmo8}
\end{equation}
From Eq.~(\ref{ecosmo1}) one sees that the relic density $f(0)$ depends essentially on the ratio $\langle v\sigma\rangle/H$. The smaller this quantity is, the less time the WIMPs can remain in thermal equilibrium and thus the larger is their relic abundance. One can show (see {\em e.g.}~\cite{Enqvist:1988dt}) that the dependence is in fact almost linear: $\Omega_{N_2} \sim H/\langle v\sigma\rangle$. With the standard expansion history of the universe $H$ is fixed, and so the characteristics of the solution are entirely dictated by the average cross section $\langle v\sigma\rangle$.  

We compute $\langle v\sigma\rangle$ in the Maxwell-Boltzmann approximation~\cite{Gondolo:1990dk}:
\begin{equation}
   \langle v \sigma \rangle = 
       \frac{1}{8m_2^{4}TK^{2}_2(\frac{m_2}{T})}
      \int_{4m_2^2}^{\infty}ds
                        \sqrt{s}(s-4m_2^2)K_1(\frac{\sqrt{s}}{T})
                        \sigma_{\rm tot}(s)
\label{ecosmo3}
\end{equation}
where $K_i(y)$s are modified Bessel functions of the second kind and $s$ is the usual Mandelstam invariant. This approximation for the collision integral is accurate to within a few per cents for massive neutrinos~\cite{Dolgov:1992wf}. For the total cross section $\sigma_{\rm tot}$ we considered the $N_2\bar{N_2}$ annihilation to the final states including all open fermion, gauge boson and scalar channels 
\begin{equation}
N_2 \bar N_2 \rightarrow f\bar{f}, \; W^{+}W^{-}, \; ZZ,\; ZH^0 
\; {\rm and} \; H^0H^0 \,.
\label{eq:channels}
\end{equation}
We omitted annihilations to  technifermions because these rates would be just a small correction to already subleading fermionic channel. Above, $H^0$ is the effective, light ``SM-like" Higgs state appearing in the mass operators (\ref{scalar_fermion}) and (\ref{scalar_fermion2}). We did not include the $SS$ final states in the scenario II, assuming that the new scalar $S$ is heavy. We computed the complete cross sections for each channel shown in~(\ref{eq:channels}) without further approximations and performed all $s$-integrals numerically. The $N_2$-gauge boson and $N_2$-Higgs couplings needed in these computations were given in section~\ref{sec:couplings}. In this work we make the assumption that the heavier, unstable neutrino $N_1$ has already decayed and is no longer present during the $N_2$ freeze-out. That is, we assume that the particle spectrum during the freeze-out is just the usual Standard Model particle spectrum and the annihilating WIMP. More details about the computation of the cross sections are given in the appendix.

It is clear that the most important parameters setting the scale of $\langle v\sigma\rangle$, and hence that of $\Omega_{N_2}$, are the WIMP mass $m_2$ and the mixing angle $\theta$. We have therefore displayed our main results as $\Omega_{N_2}$ contours in the $(m_2,\sin\theta)$-plane.  The results are only very weakly dependent on the sign of the mixing angle however, and we will always fix $\sin\theta$ positive in what follows. The charged lepton $E$ appears only as a virtual state in the $t$-channel $W^+W^-$ process and so our results depend only very weakly on $m_E$ as well. The dependence on $m_1$ can in principle be strong\footnote{Our generic system includes also the pure Dirac limit, where the predictions are of course entirely different from the ones shown here.}, but it turns out that in the region of interest, where the relic density can be large enough, and the observational constraints are satisfied (at relatively small mixing angles), also the dependence on $m_1$ 
is weak. For definiteness we have set $m_E = m_1 = 2m_2$. On the other hand, our results are very sensitive on the mass of the light composite Higgs particle $m_H$ as we shall see below. In addition to these mass parameters and mixing our results depend on the relative phase factor $\rho_{12}$. This dependence can be seen explicitly in the couplings derived in section~\ref{sec:couplings} and in the matrix elements given in the appendix. Finally, we have chosen to consider two different mass generation schemes in this paper; the first one (scenario I) using only the light composite $H^0$ and the other (scenario II), where the Majorana mass of the right-chiral state is generated by a light singlet $S$. The $N_2$-Higgs couplings in particular depend very sensitively on the choice of the scenario. 

To summarize: Our model predictions for the relic density $\Omega_{N_2}$ are most dependent on  parameters $m_2$ and $\sin\theta$. They are essentially sensitive to $m_H$, $\rho_{12}$ and on the mass generation scenario, while they have only a small subleading dependence on the masses $m_1$ and $m_E$.  

\FIGURE[t]{ \epsfig{file=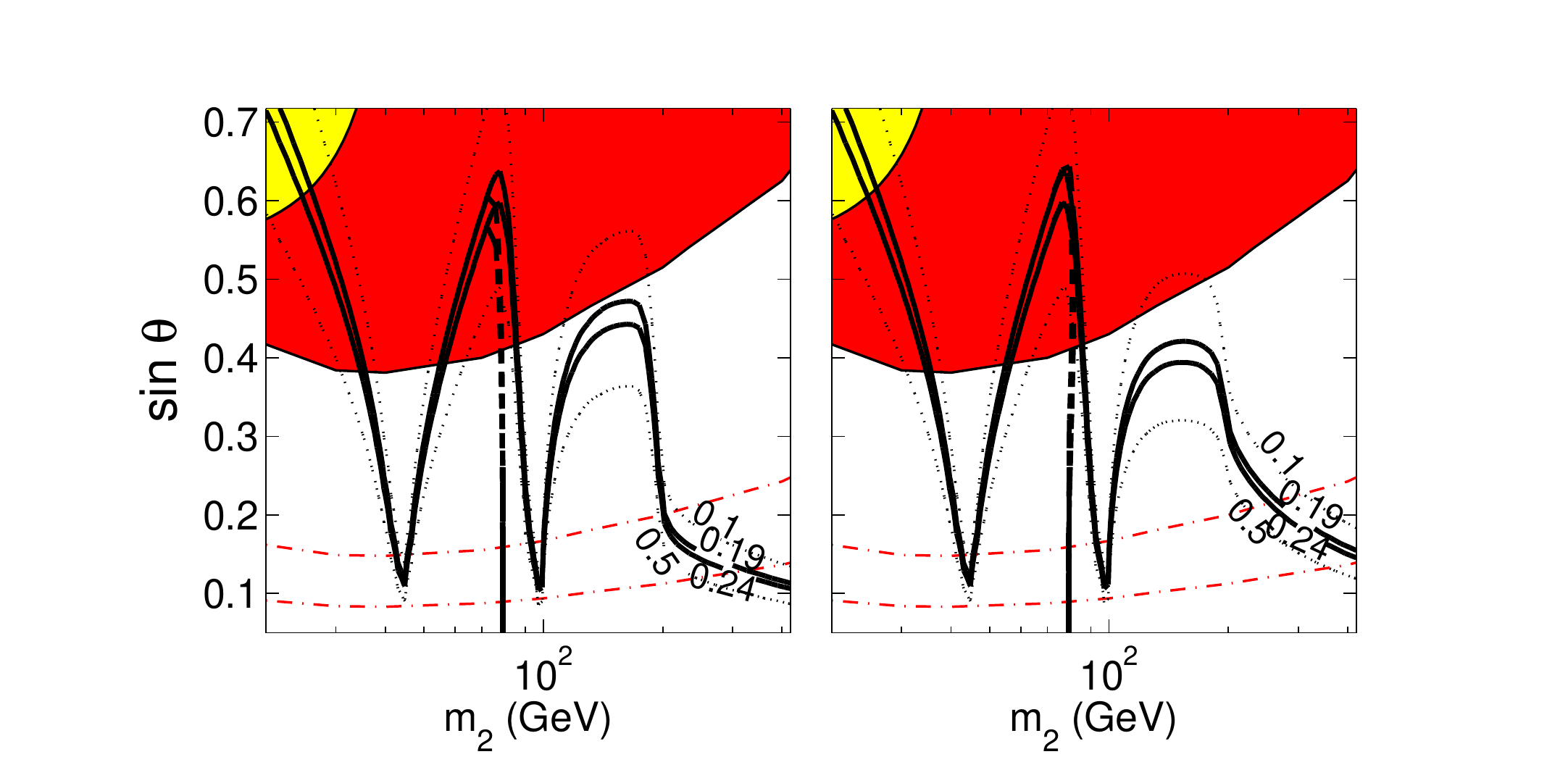,width=16cm}
\caption{Shown are constant $\Omega_{N_2}$ contours as a function of mass and mixing angle. The area between the contour lines marked by 0.19 and 0.24 is consistent with the WMAP results for the dark matter density parameter. In both panels $m_H = 200$ GeV, thick solid lines correspond to the scenario II and thick dashed lines to mass scenario I. In left panel we took $\rho_{12}=+1$ and in right panel $\rho_{12}=-1$. Thin dotted contours show additional contours for the scenario II. Yellow (light shaded) area is excluded by the LEP limits and the red (dark shaded) area is excluded by the XENON10 limits.  The red dash-dotted lines show the predicted sensitivity of the XENON100 (upper line) and XENON100 update (lower line) as given in ref.~\cite{Aprile:2009yh}.}
\label{fig:resultsH200} }
We show our results\footnote{These figures are analogous to Figs.~1-2 of ref.~\cite{Enqvist:1990yz}, with the identification $g'/g \equiv \sin\theta$. Note the typos in the powers of $g'/g$ in Eqs.~(5-7) of~\cite{Enqvist:1990yz}: they should be $\pm 4$ instead of $\pm 2$.} in Fig.~\ref{fig:resultsH200} where we used $m_H = 200$ GeV and in Fig.~\ref{fig:resultsH500} where we took $m_H = 500$ GeV. In both figures the left panel 
corresponds to the relative phase $\rho_{12}=+1$ and the right pane to $\rho_{12}=-1$. In all figures we show the contours $\Omega_{N_2} = 0.19$ and $\Omega_{N_2} = 0.23$ corresponding to the boundaries of the region where $\Omega_{N_2}$ is consistent the WMAP~\cite{Dunkley:2008ie}. The thick dashed lines mark the contours for the scenario I and thick solid lines for scenario II.  In addition we have shown by thin dashed lines the contours corresponding to $\Omega_{N_2}= 0.1$ and $\Omega_{N_2}= 0.5$ in the scenario II. We have also shown the regions of the parameters that are already excluded by the direct LEP-constraints (light yellow area) and by the XENON10 cryogenic dark matter search (dark red area), which are currently the strongest constraints on the model. We have also shown (dash-dotted curves) the sensitivity of the future XENON100 experiments on the model parameters. The observational constraints will be discussed in more detail in the next section. We finish this section by building an intuitive understanding of the relic density contours shown in Figs.~\ref{fig:resultsH200}-\ref{fig:resultsH500}. 

\FIGURE[t]{ \epsfig{file=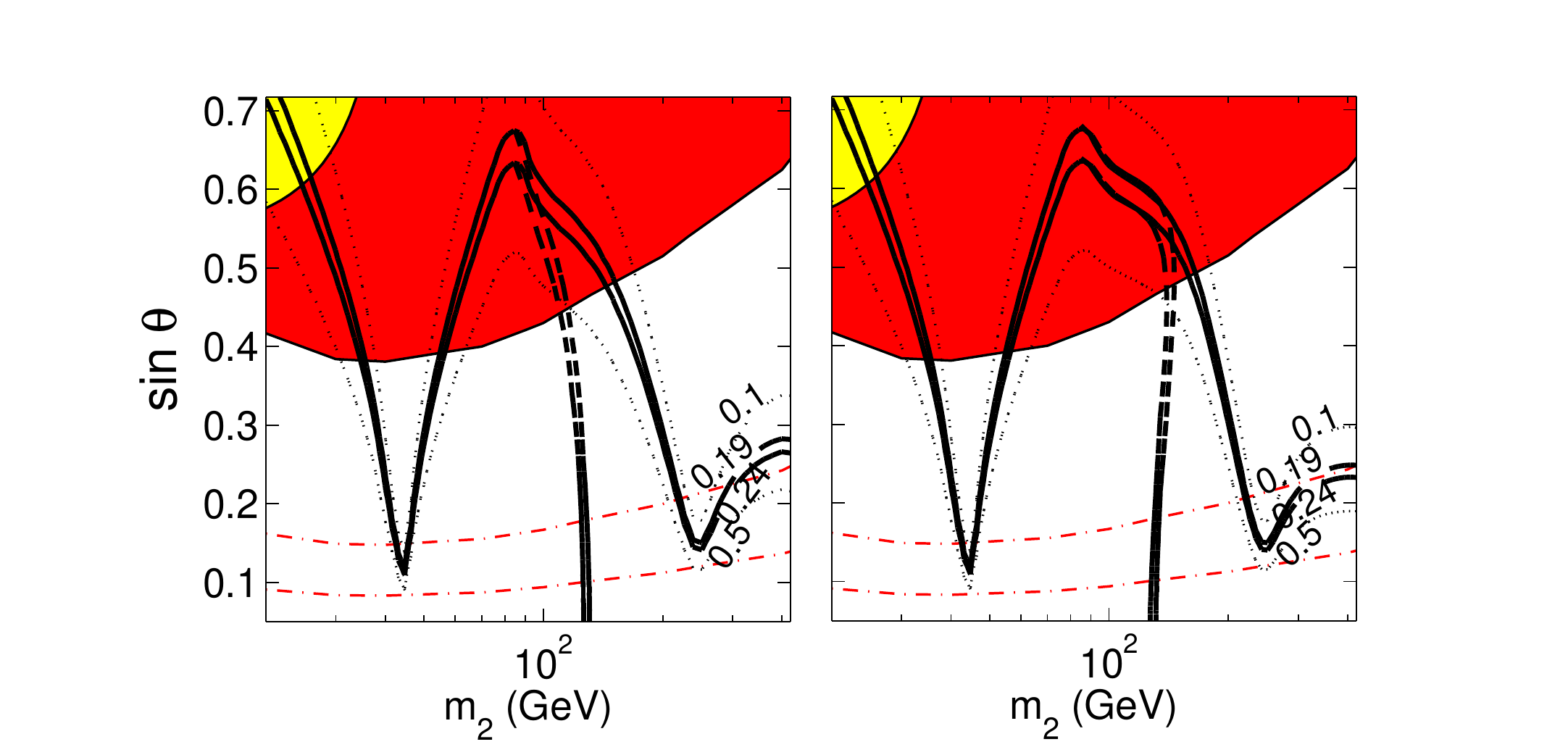,width=16cm}
\caption{Same as in figure \ref{fig:resultsH500}, but now for $m_H = 500$ GeV.}
\label{fig:resultsH500} }

Let us first concentrate on region $m_2 < M_W$, where the dominant annihilation channel is the one to standard model fermions. This cross section is simple enough to be given here explicitly:
\begin{eqnarray}
\sigma_{N_2N_2 \ra f \bar f}(s) =
  \frac{ G_F^2 m_W^4}{8 \pi s} \frac{\beta_f}{\beta_2} N_C^f
 && \left\{
   \frac{\sin^4\theta }{\cos^4\theta_W} |\hat D_Z|^2 
      g_f(\hs,m_2, \mf ) \right.
      \nonumber \\ 
 && \left. + 64 (C_{22}^h)^2 |\hat D_H|^2 \, \hs^2 \beta_2^2 \beta_f^2 \, m_2^2 \mf^2  
  \right\} \,.
\label{eq:fermionXsec}
\end{eqnarray}
Here $\theta_W$ is the Weinberg angle, $\hs \equiv s/m_W^2$, $\hat m^2_{2,f} \equiv m^2_{2,f}/m^2_W$, $\beta_X \equiv (1-4m_X^2/s)^{1/2}$ and $|\hat D_X|^2 \equiv m_W^4|D_X|^2$ for $X=Z,H$, with
\begin{equation}
D_X \equiv \frac{1}{s - m_X^2 + i \Gamma_X m_X}\,,
\label{DX}
\end{equation}
The factor $C_{22}^h$ accounts for the different coupling strengths of the lightest neutral particle to the SM-like Higgs as indicated in Table.~{\ref{c_table}}. Finally $N_C^\ell = 1$ for leptons and $N_C^q = 3$ for quarks and
\begin{equation}
g_f(\hs,m_2, \mf ) \equiv
  \Big(\sfrac{4}{3}\hs^2 - \sfrac{7}{3}\hs m_2^2 + 2m_2^2\mf^2\Big) 
  (v_f^2 + a_f^2)
+ 4 \mf^2(\hs - 6m_2^2)(v_f^2 - a_f^2)\,,
\end{equation}
where $v_f = T_{3f} - 2Q_f\sin^2\theta_W$ and $a_f = T_{3f}$ where $T_3$ is the isospin and $Q$ is the charge of the fermion. Because $\Omega_{N_2}$ is inversely proportional to the cross section, the predicted relic density is smallest near the $Z$-pole for a constant mixing angle. Conversely, to keep $\Omega_{N_2}$ constant, we have to compensate for the $Z$-pole by decreasing $\sin\theta$. This tradeoff 
%is what gives rise to 
results in the characteristic dip at $m_2 \sim M_Z/2$ apparent in Figs.~\ref{fig:resultsH200}-\ref{fig:resultsH500}. The Higgs contribution, proportional to $|D_H|^2$ in Eq.~(\ref{eq:fermionXsec}) is always subdominant to the $Z$-exchange, because of the small Yukawa couplings of the light SM-fermions. This explains why our results for different scenarios and for different choices of phase factor $\rho_{12}$ are almost identical for $m_2 \lsim M_W$.

For $M_W \lsim m_2 \lsim M_H$ the total cross section $\sigma_{\rm tot}$ is dominated by the annihilation into gauge boson final states. All channels are important quantitatively, but the most important qualitative features of the results are set by the $Z$- and $H$-mediated $s$-channel annihilations to longitudinal charged gauge bosons.  The cross section for the former of these sub-processes near the mass threshold is
\begin{equation}
  \sigma(N_2\bar N_2 \rightarrow Z^* \rightarrow W^+_LW^-_L)
        \approx \frac{G_F^2m_2^2 }{6\pi}\sin^4\theta \Big(1 - \frac{4m_2^2}{s}\Big)\,.
\end{equation}
If this was the only contribution, we would get (see {\em e.g.}~\cite{Enqvist:1990yz}.) $\langle v \sigma \rangle \approx  G_F^2m_2T\sin^4\theta/2\pi $, and eventually 
\begin{equation}
\Omega_{N_2} \approx 1.2\times 10^{-3}\frac{{\rm TeV}^2}{m^2_2 \sin^4\theta} \,.
\end{equation}
This predicts a slope $\sin\theta \approx 0.3({\rm TeV}/m)^{-1/2}$ for a constant $\Omega_{N_2}\approx 0.2$; the actual slope has a smaller coefficient because of the contribution from $ZZ$ and $ZH$ final states. This scaling was observed both in~\cite{Enqvist:1990yz} and in~\cite{Kouvaris:2007iq} and combined with the XENON10 bound it would exclude masses $m_2 \lsim 300$ GeV.  However, this simple scaling is completely changed when one includes interactions with a relatively light Higgs field. The most striking effect occurs with the scenario I where no solutions are found when $m_2 \gsim M_W$. The is due to the Higgs mediated $s$-channel annihilation to the $W^+W^-$-final state. This rate is proportional to the Higgs-WIMP coupling $C^h_{22}$, which in scenario I is not suppressed by the mixing angle. The complicated behaviour ({\em e.g.}~with the curves bending back towards smaller mass with decreasing angle) in Figs.~\ref{fig:resultsH200}-\ref{fig:resultsH500} are caused by the complex phase-angular dependence of $C^h_{22}$ and by the $m_H$-suppression in the Higgs propagator. In scenario II all Higgs couplings are proportional to the mixing and so the Higgs effects are more moderate. Nevertheless, for a relatively light Higgs mass, the above mentioned $s$-channel interaction creates the characteristic $H$-peak at $m_2 \approx m_H/2$ similar to the $Z$-peak seen at $m_2 \approx m_Z/2$. Beyond the Higgs peak the $s$-channel exchange becomes less important and one returns to the $\sin\theta \sim m^{-1/2}$-slope. For $m_2 \gsim m_H$ the behaviour changes again (only the scenario II is interesting here), because of the opening of the $HH$-final states.  In particular the four-point contact term has a large coupling, partly  because of combinatoric factors and partly because of the mass dependence $C^{h^2}_{22} \sim m_1/m_2$ at small mixing.

To summarize, we find that the superweakly interacting MWT-WIMP is a good dark matter candidate for a range of mass and mixing parameters. The preferred scenario is the one where the light neutral mass comes from a new singlet field. The WIMP of the scenario I, which uses only the field $H$ for the mass generation, is restricted to have the mass $30\,{\rm GeV} \lsim m_2 \lsim 50 $ GeV or $m_2 \sim M_W + 0.15(m_H-200)$ GeV, and $\sin\theta \lsim 0.4$. In the scenario II employing a new singlet $S$, the high mass solution is opened up to a region $m_2 \gsim 100$ GeV and $\sin\theta \lsim 0.4 - 0.5$ with exact bounds depending on the mass of the $H$-field. Let us note that the coincidence regions where the WIMP mass is close to some resonance, $m_2 \approx M_X/2$, are generic sweet spots for the superweakly interacting dark matter scenarios; in these cases DM particles can have a sufficiently large annihilation rate despite a very small coupling, which makes them difficult to detect in direct searches. Nevertheless, the near future XENON100 updates will be able to either detect, or all but rule out the WIMP proposed here, as shown by the predicted sensitivity of these searches in~Figs.~\ref{fig:resultsH200}-\ref{fig:resultsH500}.

Let conclude this section on a note on the MWT-WIMP scenario without mixing considered in~\cite{Kainulainen:2006wq}. There a sufficiently small relic density was not arranged by a superweak cross section like here, but by a large expansion rate of the universe in the context of quintessence-like dynamical dark energy model. As we have mentioned, this scenario is essentially ruled out by the XENON10 data. This is in fact evident from our  Figs.~\ref{fig:resultsH200}-\ref{fig:resultsH500}, where the line $\sin\theta \equiv 1$ would correspond to any desired $\Omega_{N_2}$-contour in the pure Majorana case of ref.~\cite{Kainulainen:2006wq}. The Dirac case is even more constrained. If we changed the expansion history of the universe in the present mixing scenario, the constant $\Omega_{N_2}$-curves would tend to move towards a larger mixing, and hence closer to or within the excluded region. In other words, changing the expansion history would change the predictions in a way that would further restrict the allowed parameter space. 

\section{Constraints}
\label{sec:cryo}

In previous section we presented our main results, including the presently strongest observational constraints. Here we will analyze in more detail how these constraints on the ($m_N,\theta$)-parameter space arise. Moreover, we will consider the present sensitivity and the future reach of the other dark matter searches, such as CDMS~\cite{Ahmed:2008eu}, Super-Kamiokande~\cite{Desai:2004pq} and IceCube~\cite{Abbasi:2009uz} experiments on the parameters of our model. Finally, we will also compute the constraints from the precision electroweak data. Here the strategy is somewhat different, because the oblique corrections are sensitive mostly to the more strongly interacting (heavier) neutrino and to the new charged lepton. Fortunately the relic density analysis depends on these parameters only very weakly, as we discussed in the previous section. Oblique corrections will therefore be computed in the $(m_1,m_E)$-plane for a number of cosmologically acceptable ($m_N,\theta$)-pairs.

\subsection{LEP-limits}

Let us start by reviewing the constraint coming from the LEP-measurement of the $Z$-boson decay width to an invisible sector~\cite{Ellis:1990}. Written in terms of the number of massless light neutrino species the allowed width to the invisible sector~\cite{Eidelman:2004wy}: 
\begin{equation}
 N_{\nu} \equiv \frac{\Gamma(Z\rightarrow \rm{inv.})}{\Gamma_{\rm{theory}}(Z\rightarrow \nu \bar{\nu})}
                 = 3.00 \pm 0.08.
\label{cosmo9}
\end{equation}
The known three light neutrino families almost saturate this bound, leaving only the deviation of the experimental value $N_{\nu}$ from three for the new invisible particles. At the $Z$-peak, the relation between the $Z$-boson decay to a massless SM-neutrino and to our heavy neutrino is $\Gamma(Z\rightarrow N_2 \bar{N_2}) = \sin^4{\theta} \beta_{2}^{3} \times\Gamma(Z\rightarrow \nu \bar{\nu})$, so that from Eqn.~(\ref{cosmo9}) we get the constraint:
\begin{equation}
 0.08 > N_{\nu}-3 = \sin^4{\theta} \times \beta_{2}^{3},
\label{ecosmo10}
\end{equation}
where $\beta_{2}$ is the neutrino velocity factor with $s = m_{Z}^{2}$. The exclusion contours corresponding to this limit were shown by the yellow shaded area in the Figs.~\ref{fig:resultsH200} and~\ref{fig:resultsH500}.

\subsection{Cryogenic and other indirect detection limits}

Adapting the constraints from a given dark matter search to a particular WIMP model is not always straightforward. The problem is that the WIMP-nucleus interactions depend on the precise form of the WIMP-nucleus and of the WIMP-nucleon couplings as well as on the proton and neutron structure functions within the target nucleus. In an ideal situation one would compute precise experiment-specific expected count rate for the WIMP model one is interested in: 
\begin{eqnarray}
\label{cosmo11}
 N & = & m_d \tau \sum_a^{\rm proc.}\sum_{i}^{\rm{isot.}} R_{ia} 
\\ \nonumber
   & = & 
   m_d \tau \sum_a^{\rm proc.}\sum_{i}^{\rm{isot.}} 
   \frac{\rho_2 Y_i}{m_2 m_{N_i}} 
   \int_{E_T}^{E_{\rm{max}}}dQ\ 2 m_{N_i}    \int_{v_{\rm{min}}(Q)}^{v_{\rm{esc}}}dv v f_1(v) 
        \frac{\sigma_{ia}}{4 \mu_i^2v^2} F_{ia}^2(Q) \chi_a(Q) \,.
\end{eqnarray}
This is a fairly complicated expression with many different parameters: $m_d$ is the mass of the detector, $\tau$ the detector exposure time, $R_i$ is the fractional count rate, $Y_i$ is the fractional abundance and $m_{N_i}$ is mass of the target nuclide of a specific isotope I and $\rho_2 = 0.3 \rm{GeV}/\rm{cm}^2$ and $m_2$ are the local WIMP density and mass. $Q$ is the energy deposited on the nuclide in the elastic collision, $E_T$ is the detector threshold and $E_{\rm{max}}$ is the maximum energy which nuclide can get from the impact. Furthermore, $v$ is the WIMP velocity, $f_1(v)$ is the WIMP velocity distribution in the earth frame~\cite{Jungman:1995df,Kainulainen:2006wq}, $v^2_{\rm{min}}(Q)= Q m_{N_i}/2 \mu_i^2$ corresponds to the minimum velocity of the WIMP, $v_{\rm{esc}}$ is its escape velocity from the galaxy gravitational potential and $\mu_i = m_2 m_{N_i}/(m_2 + m_{N_i})$ is the reduced mass of the WIMP and the target nuclide.
\TABLE[t]{
\begin{tabular}{|c|c|c|}
\hline
Factor & Spin-dependent (axial vector) & Spin-independent (scalar) \\   
       & coupling                      & coupling  \\ \cline{1-3}
$C_{i}$& $ \frac{8}{\pi}[a_p \langle S_p \rangle + a_n \langle S_n \rangle]^2 \frac{J+1}{J} $ & $ A_{i}^{2} $ \\ \cline{1-3}
$F_i^2(Q)$& $ \frac{S(q)}{S(0)} $ & $ \left[\frac{3j_1(qR_1)}{qR_1} \right]^2 \exp[-(qs)^2] $ \\ \cline{1-3}
\end{tabular}
\caption{Shown are the spin-dependent and spin-independent neutrino-nuclide coupling factors. In spin-dependent case the factors $C_{i}$ and $F_i^2(Q)$ are from~\cite{Angle:2008we} the equations (2) and (3) respectively. In spin-independent case $F_i^2(Q)$ is from~\cite{Engel:1991}. ($q \equiv \sqrt{2Qm_{N_i}}.)$)}
\label{NuclearCouplinstable} }
Finally, the cross section $\sigma_{ia}$ is given by
\begin{equation}
\sigma_{ia} \equiv C_{ia} \mu_i^2 \tilde \sigma_{0a}
\label{eq:cryoxsec}
\end{equation}
where $C_{ia}$ is an enhancement factor shown in table~\ref{NuclearCouplinstable}, and $\tilde \sigma_{0a}$ is a dark matter model dependent constant which is independent of the isotope. The formula (\ref{cosmo11}) involves also the nuclear isotope, interaction process and nuclear model specific functions $F_{ia}^2(Q)$ and the process and experimental setup dependent function $\chi_a(Q)$ that characterizes for example the detection efficiencies and cuts used to reduce the background. Unfortunately these functions are often insufficiently detailed by the experiments, making it difficult or impossible to use Eq.~(\ref{cosmo11}) directly. Conversely, it is not feasible for experiments to put forth specific constraints on all different DM-models and so, in a compromise between accuracy and generality, the observational limits are typically expressed in terms of pure WIMP-proton and WIMP-neutron cross sections. 

The most important quantity affecting the constraints comes from the nuclear spin dependence of the WIMP-nucleon interaction. Interactions which require a nucleon spin-flip to proceed, are called {\em spin-dependent} interactions. Our WIMPs are Majorana particles with an axial vector coupling to $Z$ and a scalar coupling to $H$. Thus, the $Z$-mediated interaction process for our model is spin-dependent, while the $H$-mediated process is a {\em spin-independent} one, which does not require a nucleon spin-flip to proceed. All WIMP-nucleon interactions are very soft and so the spin-dependent interactions are only seen with nuclei with (an) unpaired valence nucleon(s). Because WIMPs couple differently to protons and neutrons, the identity of these valence nucleons (whether mostly a proton or a neutron) gives rise to a further sensitivity on nuclear physics details. At any rate, to get the most accurate bound on a given model one should sum the spin-independent and spin dependent processes incoherently (because these processes have different final states) in the predicted count rate as indicated in Eq.~({\ref{cosmo11}}). However, in practice one typically assumes that WIMP has either purely spin-dependent or purely spin-independent interactions.

For our model, the best current limit comes from the cryogenic dark matter search XENON10 experiment~\cite{Angle:2008we}. Fortunately, this experiment has given their (spin-dependent) constraints not only for the standard proton and neutron interactions, but also for a standard model 4$^{\rm{th}}$ family Majorana neutrino. Better yet, reference~\cite{Angle:2008we} explicitly plots the expected count rate in their detector for this case. Now, the spin dependent count rate $N$ predicted for our WIMP through Eq.~(\ref{cosmo11}), differs from the standard model case only by a simple scaling of the cross section factor $\tilde \sigma_0$ in Eq.~(\ref{eq:cryoxsec}):
$\tilde \sigma_{0,\rm SM} \rightarrow \tilde \sigma_{0,\rm Mix} = \sin^4\theta\tilde \sigma_{0,\rm SM}$. Using this information we can convert the XENON10 results for a 4$^{\rm{th}}$ family SM-neutrino to an upper limit on the mixing angle as a function of mass:
\begin{equation}
\sin \theta (m_2)< \left(\frac{N_{\rm{limit}}(m_2)}{N_{\rm{SM}}(m_2)}\right)^{1/4}.
\label{eq:uppersin}
\end{equation}
The function $N_{\rm{SM}}(m_2)$ was read from the left panel in Fig.2 of ref.~\cite{Angle:2008we} and the function $N_{\rm{limit}}(m_2)$ was approximated by a linear interpolation between the values of $N_{\rm{SM}}(m_2)$ at the high and low mass ends of the SM-exclusion region in the same figure. The red areas in the Figs.~\ref{fig:resultsH200} and~\ref{fig:resultsH500} show the excluded region corresponding to the upper limit (\ref{eq:uppersin}). Finally, the future reach of the XENON100 updates can be simply estimated by scaling the above limiting angle (\ref{eq:uppersin}) by the fourth root of the mass-day exposure ratios of the present and future searches:
$\sin\theta_{\rm lim} \rightarrow \sin\theta_{\rm lim} (E_{Xe10}/E_{upg})^{1/4}$. We use $E_{Xe10} = 136$ kg-days for Xenon10, $E_{Xe100} = 6000$ kg-days for Xenon100 and $E_{Xe100u}= 60000$ kg-days for the Xenon100 upgrade~\cite{Aprile:2009yh}. The results are displayed by the red dash-dotted curves in Figs.~\ref{fig:resultsH200} and~\ref{fig:resultsH500}. 

\FIGURE[t]{ \epsfig{file=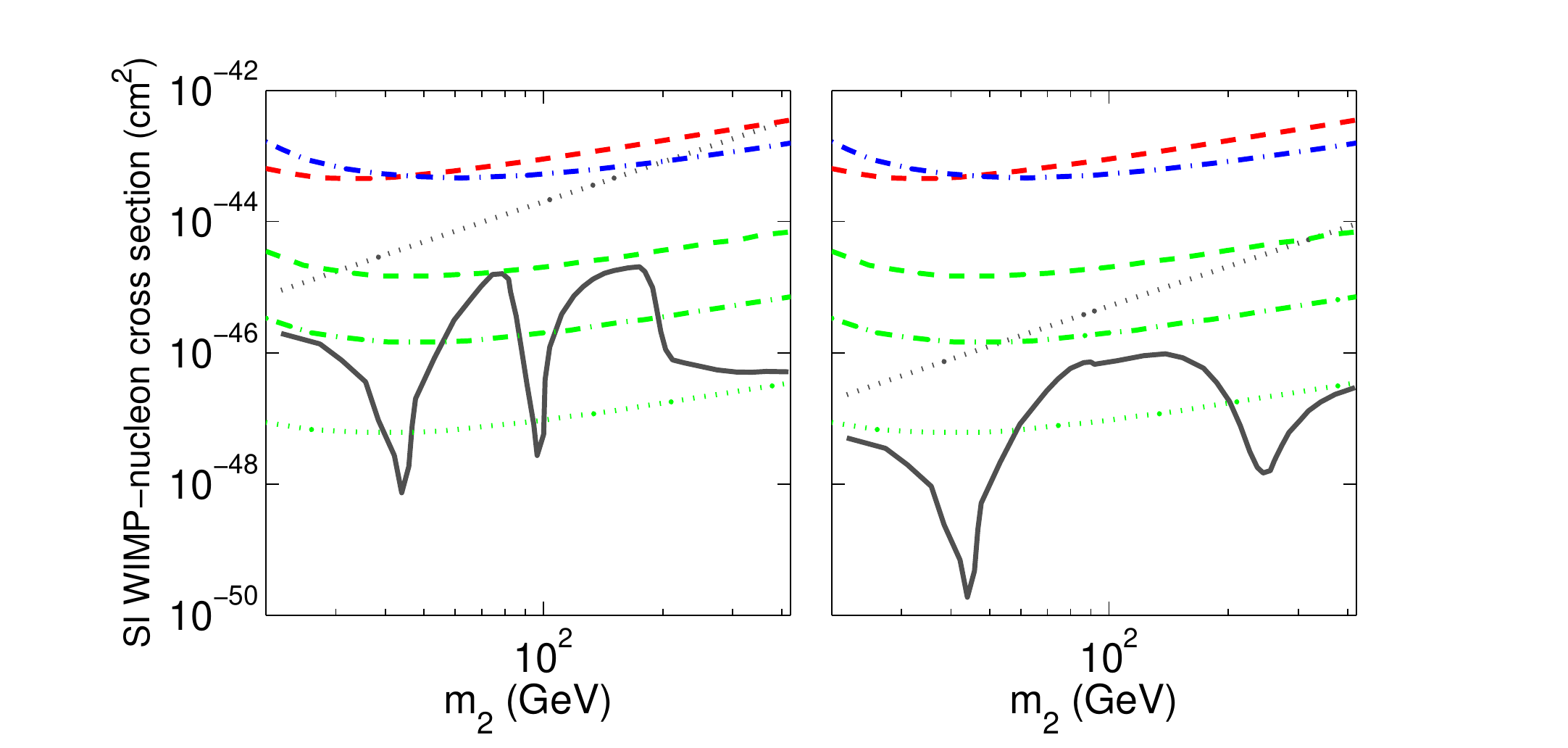,width=16cm}
\caption{Shown are spin-independent WIMP-nucleon cross sections. Dash dotted blue line shows the latest CDMS constraint~\cite{Ahmed:2008eu} and the red dashed line corresponds to XENON10 limit from~\cite{Angle:2007}. The solid black lines show the predicted cross section yielding $\Omega_{N_2}=0.214$ in scenario II with  $m_H=200$ GeV and $\rho_{12}=+1$ (left panel) and with $m_H=500$ GeV and $\rho_{12}=-1$ (right panel). The dotted gray line shows the standard, mixing angle independent $\Omega_{N_2}=0.214$-cross section for comparison. 
Finally, the light dashed (green) lines show the sensitivity of the XENON100 experiments and the light dotted (green) line that of the XENON1T experiment~\cite{Aprile:2009yh}. For the production of the experimental curves here and in fig.~\ref{fig:SpinDeLimits} we used have used the tools of ref.~\cite{dmtools}.}
\label{SpinInlimits} }

As mentioned above, our WIMP has also spin-independent interactions mediated by the Higgs field, which were not accounted for in the treatment leading to the constraint (\ref{eq:uppersin}). Let us check that the limit coming from this process is really subdominant. This is not immediately clear, because the spin-independent channel is coherent, and it is thus enhanced with respect to a spin-dependent one by a factor $\sim A^2 F^2(Q)$ where $A$ is the mass number of the target nucleus and $F^2(Q)$ is a shadowing correction shown in table~\ref{NuclearCouplinstable}; for the XENON10 experiment the enhancement is roughly $\sim 10^4$. The spin-independent limit from XENON10 is given in~\cite{Angle:2007} in terms of the WIMP-nucleon cross section. In our model this cross section (in the zero momentum transfer limit\footnote{The zero momentum transfer approximation could be improved by the methods of ref.~\cite{Gondolo:1996qw}. These corrections are quite small however, and they would not affect the final conclusion.}) is given by
\begin{equation}
 \sigma_{0}^{\rm n} = \frac{G_F^2\mu_{\rm n}^2}{4 \pi} \frac{m_2^2 m_{\rm n}^2}{m_H^4}  \, (C_{22}^h)^2,
\label{cosmo14}
\end{equation}
where n refers to a nucleon and $\mu_{\rm n}$ is the WIMP-nucleon reduced mass. Note that without the $C_{22}^h$-factor the cross section (\ref{cosmo14}) would coincide with the SM Higgs-neutrino scattering cross section. Our results are shown in Fig.~\ref{SpinInlimits}. The red dashed line corresponds to the lower limit for the WIMP-nucleon cross section from XENON10, and the blue dashed line shows the same constraint from the CDMS experiment~\cite{Ahmed:2008eu}. The solid line shows WIMP-nucleon cross-section of Eq.~(\ref{cosmo14}) corresponding to the parameters $m_2$ and $\sin\theta(m_2)$, constrained to give the solution $\Omega_{N_2} = 0.214$, shown in Figs.~\ref{fig:resultsH200}-\ref{fig:resultsH500}. We show the results only for the scenario II, where $C_{22}^h = \sin^2\theta$ (see table~\ref{c_table}). The predicted cross sections in all cases shown are well below the observational limit showing that the spin-independent constraints are clearly weaker than the spin-dependent ones despite the coherence enhancement.  

\FIGURE[t]{ \epsfig{file=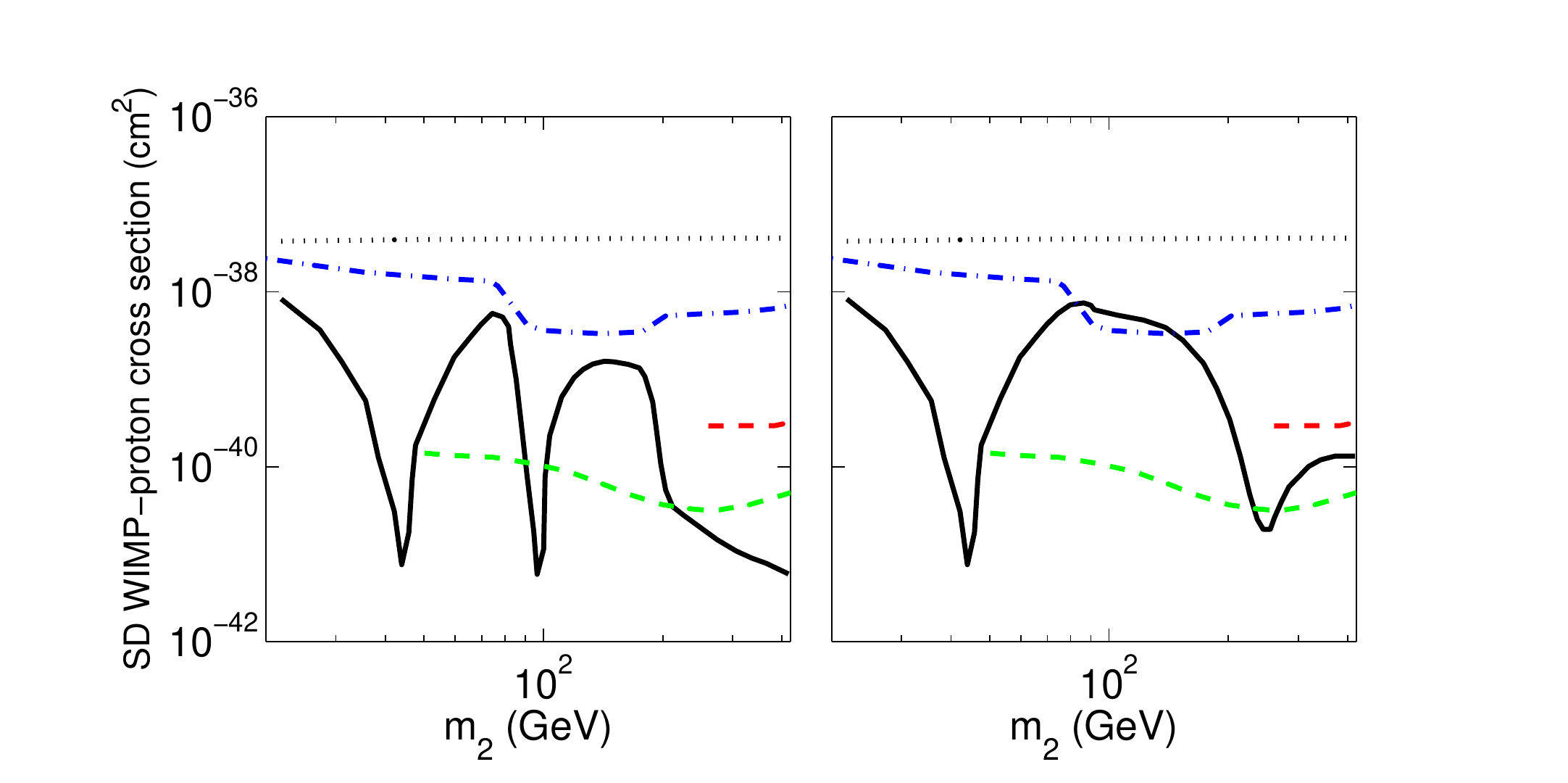,width=16cm}
\caption{Shown are spin-dependent WIMP-proton cross sections. Dash dotted blue line shows the Super-Kamiokande sensitivity. Red dashed line corresponds to the present IceCube (hard $W^+W^-$ channel) limit. Black solid lines show the predicted cross sections corresponding to $\Omega_{N_2} = 0.214$ in scenario II for $m_H=200$ GeV and $\rho_{12}=+1$ (left panel) and for $m_H=500$ GeV and $\rho_{12}=-1$ (right panel).  Again, the standard, no mixing Majorana cross section is shown for comparison with the dotted gray line. Finally, the green (light) dashed line shows the predicted IceCube DeepCore extension sensitivity~\cite{Resconi:2008fe,Abbasi:2009uz,Wikstrom:2009kw}.}
\label{fig:SpinDeLimits}}

To conclude this section we show the constraints coming from the indirect dark matter searches in Super-Kamiokande and IceCube experiments. These limits correspond to spin-dependent interactions and are expressed in terms of the standard WIMP-proton interactions. From Eqn.~(\ref{eq:cryoxsec}) and Table(\ref{NuclearCouplinstable}) we get the necessary spin-dependent WIMP-proton cross section in our model:
\begin{equation}
 \sigma_{0}^{p} = \tilde{\sigma_0} C_{p} \mu_p^2
                = \sin^4{\theta} \frac{8 G_{F}^2}{\pi} \mu_p^2 [a_p \langle S_p \rangle]^2 \frac{J+1}{J}.
\label{cosmo12b}
\end{equation}
In our calculations we have used the EMC measurement value for WIMP nucleon spin factor $a_p^2 = 0.46$ and of course $\langle S_p \rangle = 0.5$ and $J=1/2$. In Fig.(\ref{fig:SpinDeLimits}) we show the the predicted WIMP-proton cross section in our model together with the constraints as a function of the WIMP mass $m_2$. For each value of $m_2$ the mixing angle $\sin\theta$ is again chosen to give the correct dark matter density parameter value $\Omega_{N_2} = 0.214$. We have plotted the cross sections for $m_H=200$ GeV and $\rho_{12}=+1$ and for $m_H = 500$ GeV and $\rho_{12}=-1$ in the scenario II. The other cases shown in Figs.~\ref{fig:resultsH200}-\ref{fig:resultsH500} are somewhat less constrained. Although the Super-Kamiokande constraint, shown by the blue dash-dotted line, does cut the allowed parameter space, the limit is weaker than that coming from the XENON10 experiment. The constraint coming from the IceCube experiment, shown by the red dashed (short) line, is particularly interesting, because it is much more stringent than the XENON10 limit in the high mass region. Yet our model escapes the current IceCube constraint, essentially due to the light Higgs final state interactions which force a large suppression on the acceptable mixing angles (see section~\ref{sec:omega}).  However, with its projected sensitivity after the planned DeepCore extension~\cite{Resconi:2008fe,Abbasi:2009uz}, the IceCube experiment will be able to probe a large fraction the parameter space available in our model.  This is shown by the green dashed (long) line in the plot.

\subsection{Oblique constraints}
\label{sec:resultsoblique}

As we stated already in Sec. \ref{sec:oblique}, the required formulas for the evaluation of the oblique corrections can be found from \cite{Antipin:2009ks} and we will not rewrite them here but simply show the relevant results. For a fixed phase $\rho_{12}$, we consider only the parameter values $(m_2,\sin\theta)$ constrained to satisfy $\Omega_{N_2}(m_2,\sin\theta)\approx 0.214$. Then, for any given such pair of values, we evaluate the resulting constraints on the remaining mass parameters $m_1$ and $m_E$ when the precision parameters $S$ and $T$ are required to lie within limits compatible with present observations. This approach makes sense because, as we have explained above, $\Omega_{N_2}$ depends only very weakly on $m_1$ and $m_E$. We will use a rather liberal estimate $\vert S\vert\le 0.2$ and $0\ge T\ge 0.5$. Since the neutrino masses originate from non-renormalizable operators, some of the vacuum polarizations required for the calculation of oblique corrections depend on the renormalization scale. This scale dependence affects only the polarizations where we have neutrinos with masses $m_a$ and $m_b$ ($a,b=1,2$) in the loops. We have chosen the scale as $1.5m_1$ as discussed in \cite{Antipin:2009ks}. We also require the masses to satisfy $m_2\le m_1,m_E$.   

In figure \ref{fig:ST_constraints} we show the constraints from oblique corrections for three different values of $(m_2,\sin\theta)$ which were chosen to probe both small and large mixing and light and heavier values of $m_2$. The results corresponding to $\rho_{12}=+1$ and $\rho_{12}=-1$ are shown separately in left and right panels of the Figure, respectively. From these figures we see that the oblique corrections do provide nontrivial constraints on the spectra of the fourth generation leptons if the heavy neutrino is required to be responsible for the dark matter abundance of the universe. This is so since the oblique corrections are sensitive to all masses in the leptonic sector. Moreover there is sensitivity to the mixing angle, because the WIMP $N_2$ contributes with a coupling proportional to $\sin\theta$ and the heavier mass eigenstate $N_1$ with a coupling proportional to $\sim \cos\theta$. On the other hand we observe that there is still  substantial room in the parameter space for these masses. In particular the charged lepton can be relatively heavy. However, a clear pattern relating the mass of the charged lepton and the heavier neutrino eigenstate linearly can be observed. As a rough estimate for our case, $m_E\sim 2m_1$ allowing for a mild dependence of the slope on the mixing angle\footnote{Note that in section~\ref{sec:omega} we used $m_E = m_1$. However, we checked that changing the mass hierarchy to $m_E = 2m_1$ has little effect on $\Omega_{N_2}$, as expected.}. This result is compatible with the corresponding results in \cite{Antipin:2009ks}.  Also, the results for positive and negative $\rho_{12}$ are qualitatively similar. The hierarchy $m_2\le m_1,m_E$ which we have imposed is easy to observe from the figures. Also, since the coupling of the charged lepton to electroweak currents is proportional to $\cos\theta$, we would expect that for small values of $\sin\theta$ the oblique corrections would favor heavy charged lepton due to large value of the coupling of this state; this feature is observed in the above figures and appears more pronounced for the case $\rho_{12}=-1$. Also, due to the interplay of the angles, the most accessible (in view of the colliders) mass ranges seem to correspond to $\sin\theta\sim 1/2$ which ``optimizes'' the couplings of all states to be of similar magnitude. 

\FIGURE[t]{\epsfig{file=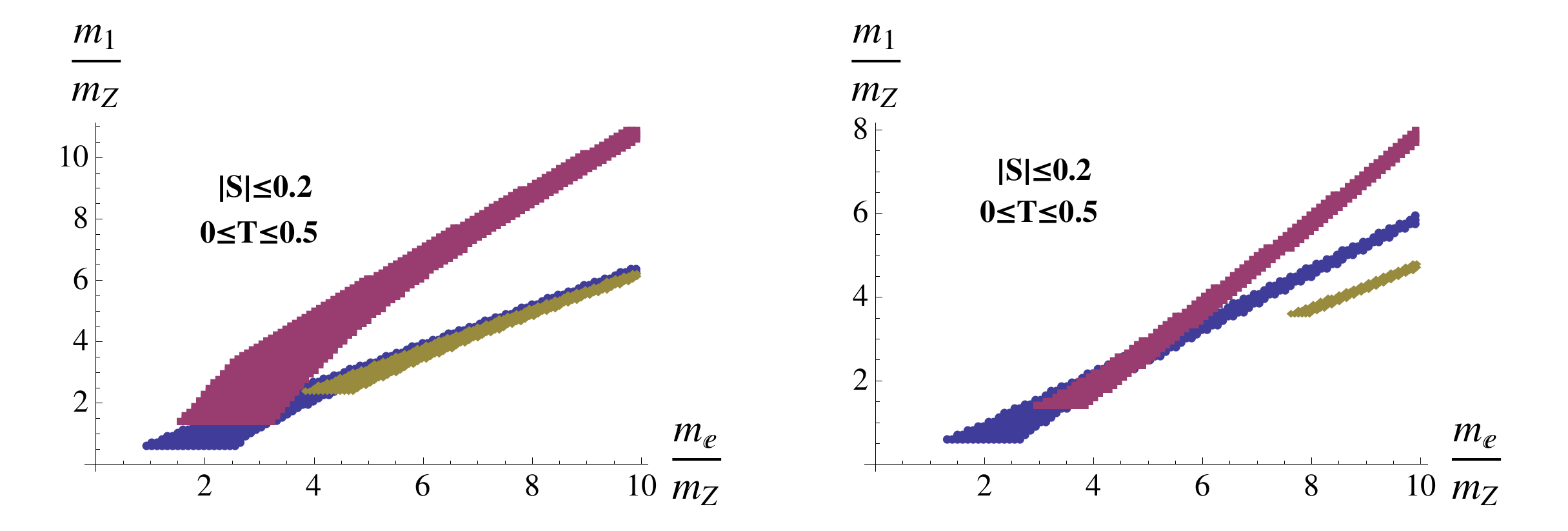,width=15cm} 
\caption{Allowed values of $m_1$ and $m_E$ as constraints $\vert S\vert\le 0.2$ and $0\ge T\ge 0.5$ are imposed in addition to the constraint $\Omega_{N_2}(m_2,\sin\theta)$. In left panel $\rho_{12}=+1$ and the three shaded domains correspond to $(\sin\theta,m_2/m_Z)=(0.282, 0.55)$ (lowermost) $(0.55,1.29)$ (top) and $(0.223,2.19)$ (middle). In the right panel $\rho_{12}=-1$ and the three shaded domains correspond to 
$(\sin\theta,m_2/m_Z)=(0.282, 0.55)$ (lowermost) $(0.59,1.29)$ (top) and $(0.207,3.28)$ (middle).}
\label{fig:ST_constraints}}

\section{Conclusions} 
\label{sec:checkout}
 
We have considered a possibility that the dark matter observed in the universe is a massive superweakly interacting Majorana neutrino suggested by recently discovered technicolor models for the electroweak symmetry breaking~\cite{Sannino:2003xe, Sannino:2004qp}. The spectrum of new particles predicted by the model contain a new massive 4th family lepton doublet and a singlet sterile state. Our WIMP is the lighter eigenstate in the mixture of the doublet neutrino and the sterile state, made stable by a discrete $Z_2$ symmetry under the exchange of any DM-sector fields. The mass scale of the new particle spectrum is predicted to be of the order of a few hundred GeV by the constraints coming from electroweak precision experiments. The relic density of such a neutrino can fall to the observationally determined range $\Omega_{N_2} \approx 0.2$ for a wide range of model parameters, the most important of which are the WIMP and the lightest Higgs masses $m_2$ and $m_H$ and the active sterile neutrino mixing angle $\theta$. Results are also sensitive on the chosen mass generation scenario, of which we have considered two distinct possibilities here. In the first scenario all masses are created using only the light doublet Higgs composite, while in the second scenario the $R$-chiral mass is created by a new singlet scalar field. The second scenario allows a much larger range of WIMP masses. Yet another parameter in our model is the relative phase $\rho_1\rho_2$ between the mass eigenstates (where $\rho_i$ are the phases necessary to make the masses positive). We pointed out that the usual Majorana Feynman rules have to be generalized to compute the necessary matrix elements in the presence of such nontrivial mixing phases. The accepted parameter region forms a thin surface $\theta = \theta(m_2,m_H,\rho_{12})$, of which we have shown several examples in the $(m_2,\sin\theta)$-plane. 

We also computed the constraints on our model parameters coming from the LEP-limits on invisible $Z$-decays, from the corrections to the oblique parameters and from direct and indirect dark matter searches. The most stringent constraint on the WIMP mass $m_2$ and mixing angle $\theta$ currently come from the cryogenic XENON10 experiment, which already excludes parts of the acceptable parameter space. We also computed the indirect bounds coming from Super-Kamiokande and IceCube, finding that the projected DeepCore extension of the IceCube experiment will be able to probe a large part of the parameter space of our model. However, the most stringent future constraints on the model parameters will be set by the upcoming XENON100 and XENON1T experiments. Further to these limits, the oblique constraints were shown to be satisfied within narrow strips in the plane ($m_E,m_1$) of the heavy lepton $E$ and the heavier neutral state $N_1$ masses such that the exact position and the area of the strip depends on the given acceptable ($m_2, \sin \theta$)-pair. Currently it appears that the direct and indirect DM-searches will have the better reach to the viability of the model than does the precision electroweak data. In particular the forthcoming XENON100 updates will have the sensitivity to rule out almost the entire parameter space of our model, or in the more favourable case to detect a WIMP with the characteristics proposed by the Minimal Walking Technicolor model.

%%%%%%%%%%%%%%%%%%%%%%%%%%%%%%%%%%%%%%%%%%%%%%%%%%%%%%%%%%%%%%%%%%%%%%%%%%%%%

\section*{Acknowledgments}
\noindent We thank Jouni Suhonen for enlighting discussions and Alfredo Ferella for correspondence concerning the results of the XENON10 experiment. JV thanks the Finnish Academy of Sciences and Letters, Vilho, Yrj\"o and Kalle V\"ais\"al\"a foundation, and the Academy of Finland, Graduate School in Particle and Nuclear Physics (Graspanp) for grants.

%%%%%%%%%%%%%%%%%%%%%%%%%%%%%%%%%%%%%%%%%%%%%%%%%%%%%%%%%%%%%%%%%%%%%%%%%%%%%

\section{Appendix: Cross sections and unitarity}
%\label{sec:appendix}

Computing Majorana cross sections is much more complicated than finding the corresponding quantities for Dirac particles. Moreover, as shown in section \ref{sect:rho-business}, the Majorana phases enter into the cross sections in quite a nontrivial way with mixing fields. We therefore believe that it is worthwhile to present some of the details of our calculations here. While most cross sections used in this paper are too long to be written down explicitly, their matrix elements can be expressed in a fairly compact form. Even computing these matrix elements requires lengthy reductions and combination of many independent contractions, and from these results one can already appreciate the nontrivial effects of the mixing phases. Matrix elements may also be used as a starting point for other problems involving these processes. According to our definition (\ref{Neigen}) $N^c = \rho N$. This means that the phases $\rho_i$ must appear in the field operators, which now become
\begin{equation}
N_i = \sum_h \int \frac{{\rm d}^3k}{(2\pi)^3} 
\left[ \hat a_{h{\bf k}i} u_{hi}(k) e^{-ikx} + \rho_{i} \hat a_{h{\bf k}i}^\dagger v_{hi}(k) e^{ikx}\right] \,,
\end{equation}
where the sum is over the helicities $h$. This operator gives rise to the  following nontrivial contractions in the momentum space Feynman rules:
\begin{eqnarray}
\bcontraction{}{N}{}{a} N_i \hat a_{h{\bf p}i}^\dagger = u_{hi}(p) 
\quad &{\rm and}& \quad
\bcontraction{}{N}{}{a} {\bar N_i} \hat a_{h{\bf p}i}^\dagger 
= \rho_{i}^* \bar v_{hi}(p)
\nn\\
\bcontraction{}{a}{{}_{h{\bf p}i}}{N} \hat a_{h{\bf p}i} N_i  = \rho_iv_{hi}(p) 
\quad &{\rm and}& \quad
\bcontraction{}{a}{{}_{h{\bf p}i}}{N} \hat a_{h{\bf p}i} {\bar N_i} 
= \bar u_{hi}(p) \,.
\label{ucontract}
\end{eqnarray}
Only the first two of these will be needed in our computations because in all our processes neutrinos appear in the initial state only. In Majorana case one also has four different internal contractions, or propagators:
\begin{eqnarray}
\bcontraction{}{N}{(x)_i}{N} N_i(x){\bar N_i}(y) 
&=& iS_i(x-y) \nn\\
\bcontraction{}{N}{(x)_i}{N} N_i(x) N_i(y) 
&=& iS_i(x-y)  (-\rho_i C) \nn\\
\bcontraction{}{N}{(x)_i}{N} {\bar N_i}(x) {\bar N_i}(y) 
&=& (\rho^*_i C) iS_i(x-y) \nn\\
\bcontraction{}{N}{(x)_i}{N} {\bar N_i}(x) N_i(y) &=& 
(\rho^*_i C) iS_i(x-y) (-\rho_i C) \,,
\label{propcontract}
\end{eqnarray}
where $iS_i(x-y)$ is the standard Dirac propagator for the mass eigenstate I, and $C$ is the usual charge conjugation matrix. The recipe for the cross section calculations is very simple. One writes down the $S$-matrix element in the standard way, forms all possible contractions and uses Eqs.~(\ref{ucontract}) and (\ref{propcontract}) to reduce these contractions in terms of Dirac matrices. Although a great number of independent contractions are generated in this way, the final results of the combinatorics results in rather simple expressions for the matrix elements. One then might expect that there should be an effective set of Feynman rules that would allow reducing the amount of combinatorics, and several papers have indeed put forth such rules, see for example~\cite{Denner:1992me}. Unfortunately none of the existing papers offers rules that could be used in the case of mixing fields with nontrivial phase factors $\rho_i\rho_j \neq 1$.  We believe that simplified rules can be expressed also for mixing fields, but this work will be pursued elsewhere~\cite{ourMajrules}.\\
%\subsection{Gauge boson final states}

For the Gauge boson final states $N_2N_2\rightarrow CC$, where $C=W,Z$, 
the matrix elements can be generically written in the form
\begin{equation}
{\cal M}_{CC} = \frac{-ig^2}{2} \epsilon^C_{\mu_1}(k_1)\epsilon^C_{\mu_2}(k_2) 
\; \bar v_{N_2}(p_1) \rho_1 \Gamma^{\mu_1\mu_2}_{CC}\, u_{N_2}(p_2) \,,
\label{MatrixCC}
\end{equation}
where $\epsilon^C_\mu$ is the gauge boson polarization vector.
The reaction $N_2N_2\rightarrow WW$ is mediated by $t$- and $u$-channel $E$-lepton exchanges and $s$-channel $Z$- and Higgs boson exchanges. In each channels one has two independent Majorana contractions.  Combining all contractions the final matrix element can be compactly written in the form (\ref{MatrixCC}), with
\begin{eqnarray}
\Gamma^{\mu_1\mu_2}_{WW} &=&  
     \sin^2\theta \left( 
     D_{t}^L  \gamma^5\gamma^{\mu_1}(\ksl_1-\psl_1) \gamma^{\mu_2}
   + D_{u}^L  \gamma^5\gamma^{\mu_2}(\ksl_2-\psl_1) \gamma^{\mu_1}
      \right)
\nonumber\\ [0.2 cm] 
   &+& \sin^2\theta  D_{Z}  \gamma^5 \gamma_{\lambda} 
   V^{\lambda \mu_1 \mu_2}_{k_1+k_2,-k_1,-k_2} 
   \, - \, 2 m_2 C_{22}^h D_H g^{\mu_1\mu_2} \,,
\label{GammaWW}
\end{eqnarray}
where the three-gauge-boson coupling factor is
\begin{equation}
V^{\lambda \mu_1 \mu_2}_{k_1+k_2,-k_1,-k_2} 
  = (k_2-k_1)^\lambda g^{\mu_1\mu_2}
   -(k_1+2k_2)^{\mu_2} g^{\lambda\mu_1}
   +(k_2+2k_1)^{\mu_1} g^{\mu_2\lambda} \,
\end{equation}
and the propagator factors $D_X$ are given by (\ref{DX}) for $X = Z,H$ and
\begin{equation}
D_a^L \equiv \frac{1}{a + m_L^2} \,,
\label{eq:simpleprop}
\end{equation}
for  $a = t,u$.  Like fermion final state processes, the $WW$-process do not involve neutrino mixing vertices, and so the result involves no nontrivial phases and it agrees with the one obtained by use of the standard Majorana Feynman rules. Obviously, the expression for the total cross section computed from ${\cal M}_{WW}$ must be a very lengthy, and we shall not present it here. (For an analogous expression of an annihilation of Dirac neutrinos see ref.~\cite{Enqvist:1988we}.)  Let us however consider explicitly the limit $s \ra \infty$. The leading term of the total cross section now becomes:
\begin{equation}
\sigma_{N_2N_2 \ra W^+W^-}(s\gg m_X) \approx
\frac{G_F^2m_1^2}{4 \pi} (C_{22}^h - \sin^2\theta)^2 + {\cal O}\left(\frac{1}{s}\log s\right)
\,.
\end{equation}
Obviously this cross section is unitary in the scenario II but not in scanario I. The unitarity of scenario II is actually an exception and {\em nonunitarity} is a common feature for all our gauge and Higgs boson final states with mixing Majorana fields. However, the nonunitary is not a problem here, because our low-energy theory giving rise to the masses is only an effective one, known to be broken at the $\sim$TeV region.

In the process $N_2N_2\rightarrow ZZ$ one has $N_1$- and $N_2$-mediated mediated $t$- and $u$-channel processes an $H$-mediated $s$-channel processes. Here each $t-$ and $u-$ channel process consists of four independent contractions. Including two nontrivial $s$-channel contractions, one has to work out and combine 18 independent matrix elements. The final result can again be written in the form (\ref{MatrixCC}), with
\begin{eqnarray}
\Gamma^{\mu_1\mu_2}_{ZZ} &=&  
  \frac{1}{2\cos ^2\theta_W} 
   \left\{ 
     \sin^4\theta
     \left[ D_{t}^1 \gamma^{\mu_1} (\psl_1-\ksl_1+m_2) \gamma^{\mu_2}
          + D_{u}^1 \gamma^{\mu_2} (\psl_1-\ksl_2+m_2) \gamma^{\mu_1} 
      \right]
   \right. \nonumber \\   
 &&\left.  \phantom{hi} +\sin^2\theta\cos^2\theta
     \left[ D_{t}^2 \gamma^{\mu_1} (\psl_1-\ksl_1+ \tilde m_1)\gamma^{\mu_2}
          + D_{u}^2 \gamma^{\mu_2} (\psl_1-\ksl_2+ \tilde m_1)\gamma^{\mu_1}      
     \right]
   \right. \nonumber \\ 
 &&\left. \phantom{hi} +  8m_2 C_{22}^h D_H  g^{\mu_1\mu_2} 
   \right\} \,. 
 \end{eqnarray}
where $\tilde m_1 \equiv \rho_{12} m_1$. Depending on the sign of the relative phase $\rho_{12}\equiv \rho_1\rho_2$ the effective mass of the heavier field $N_2$ can be either positive or negative in this expression. Different choices for this phase lead to physically different cross sections and eventually very different predictions for the relic density. To our knowledge this is the first time when Majorana phases have been observed to have such a large effect on physical observables. In the limit $s\rightarrow \infty$ the cross section for process $N_2N_2 \ra ZZ$ in the scenario II becomes:
\begin{eqnarray}
\sigma_{N_2N_2 \ra ZZ}(s\gg m_X ) &\approx& 
  \frac{G_F^2}{8\pi }\sin^4\theta\cos^4\theta  \,(m_2^2-\rho_{12}m_1^2)^2 
 \nonumber\\
 &=& \frac{G_F^2}{8\pi } \frac{4m_D^4}{(M_L-M_R)^2 + 4m_D^2} \,.
\end{eqnarray}
Again the nonunitary behaviour is manifest. However, unitarity is again restored in the limit of no mixing, or equivalently, for a vanishing Dirac mass term. This restoration of unitarity at zero mixing is a generic result for all cross sections in the scenario II, following from the fact that in this limit the WIMP becomes a purely sterile state, which is connected to the dangerous gauge boson degrees of freedom only through the Dirac mass term. \\

The annhilation to higgs bosons, $N_2N_2 \rightarrow HH$ processes through both $N_1$ and $N_2$-mediated $t$- and $u$-channel diagrams as well as an $s$-channel Higgs exchange and the  $hhN_2N_2$-contact interaction.  After some algebra the matrix element can be written as 
\begin{equation}
{\cal M}_{HH} = \frac{-ig^2}{2} \frac{m_2^2}{M_W^2}\bar v_{N_2}(p_1) \rho_1 A_{HH} \, u_{N_2}(p_2) \,,
\label{MatrixHH}
\end{equation}
where
\begin{eqnarray}
A_{HH} &=& 2(C_{22}^h)^2 
     \left[D^2_{t}(2m_2 - \ksl_2) + D^2_u(2m_2 - \ksl_1) \right]
\nonumber \\  
&+&   \sfrac{1}{2} (C_{12}^h)^2
        \left[D^1_{t}(m_2 + \tilde m_1- \ksl_2) 
            + D^1_{u}(m_2 + \tilde m_1 - \ksl_1) \right]
    \nonumber \\   
&+& 3 C_{22}^h \frac{m_H^2}{m_2} D_H + \frac{2}{m_2} C_{22}^{h^2} \,.
\label{HiggsCC}
\end{eqnarray}
where again $\tilde m_1 \equiv \rho_{12} m_1$. The coefficients $C^h_{22}$, $C^h_{12}$ and $C^{h^2}_{12}$ are given in the Table~\ref{c_table} and the propagators $D_a$ in Eq.~(\ref{eq:simpleprop}).

Finally, the annihilations to the mixed final state $ZH$ are mediated by the $t$- and $u$-channel $N_1$ and $N_2$ exchanges, as well as by an $s$-channel $Z$-exchange diagram. We find:
\begin{equation}
{\cal M}_{ZH} = \frac{-ig^2}{2\cos^2 \theta_W} \frac{m_2}{M_Z}  \epsilon^Z_\mu(k_1) \;
\bar v_{N_2}(p_1) \, \rho_1 A_{HZ}^\mu \, u_{N_2}(p_2) \,,
\label{MatrixHH2}
\end{equation}
where
\begin{eqnarray}
A_{ZH}^\mu &=& \sin^2\theta C_{22}^h 
     \left[D^2_{t}\gamma^\mu \gamma^5(\psl_2 - \ksl_2 + m_2) 
         + D^2_{u}(\psl_2 - \ksl_1 + m_2)\gamma^\mu \gamma^5 \right]
\nonumber \\  
&+&  \sfrac{1}{4}\sin 2\theta \rho_{12}C_{12}^h
     \left[D^1_{t}\gamma^\mu \gamma^5(\psl_2 - \ksl_2 + \tilde m_1) 
         + D^1_{u}(\psl_2 - \ksl_1 + \tilde m_1)\gamma^\mu \gamma^5 \right]
\nonumber \\  
&+&  \frac{2M_Z^2}{m_2}\sin^2 \theta D_Z\gamma^\mu \gamma^5 \,.
\label{HiggsCC2}
\end{eqnarray}
When matrix elements are given in these compact form, it is a straightforward to compute the cross sections using some algebraic manipulation program. We reduced all the traces using FEYNCALC, after which the resulting cross sections were transported to a dedicated fortran code which evaluated $\langle v\sigma \rangle$ at a given temperature $T$ from the integral expression (\ref{ecosmo3}).

%%%%%%%%%%%%%%%%%%%%%%%%%%%%%%%%%%%%%%%%%%%%%%%%%%%%%%%%%%%%%%%%%%%%%%%%%%%%%

\end{document}